\documentclass[10pt, conference]{IEEEtran}
\IEEEoverridecommandlockouts
\usepackage{etex}
\usepackage{subfigure}
\usepackage{graphicx}
\usepackage{amsmath}
\usepackage{bm}
\usepackage{url}
\usepackage{stmaryrd}
\usepackage{balance}
\usepackage{amssymb}
\usepackage{pifont}
\usepackage[usenames,dvipsnames]{pstricks}
\usepackage{epsfig}
\usepackage{epstopdf}
\usepackage{multirow}
\usepackage{booktabs}
\usepackage{pgffor}
\usepackage{tikz}
\usepackage{algorithm,algorithmic}
\usepackage{cite}
\newcommand{\pro}{dpMood}

\def\BibTeX{{\rm B\kern-.05em{\sc i\kern-.025em b}\kern-.08em
    T\kern-.1667em\lower.7ex\hbox{E}\kern-.125emX}}

\begin{document}
\title{dpMood: Exploiting Local and Periodic Typing Dynamics for Personalized Mood Prediction}

\author{
\IEEEauthorblockN{
He Huang\IEEEauthorrefmark{1},
Bokai Cao\IEEEauthorrefmark{2},
Philip S. Yu\IEEEauthorrefmark{1},
Chang-Dong Wang\IEEEauthorrefmark{3},
Alex D. Leow\IEEEauthorrefmark{4}
}
\IEEEauthorblockA{\IEEEauthorrefmark{1}
Department of Computer Science, University of Illinois at Chicago, IL, USA;
\{hehuang, psyu\}@uic.edu}
\IEEEauthorblockA{\IEEEauthorrefmark{2}
Facebook Inc., Menlo Park, CA, USA;
caobokai@fb.com}
\IEEEauthorblockA{\IEEEauthorrefmark{3}
School of Data and Computer Science, Sun Yat-sen University, Guangzhou, China;
changdongwang@hotmail.com}
\IEEEauthorblockA{\IEEEauthorrefmark{4}
Departments of Psychiatry and Bioengineering, University of Illinois at Chicago, IL, USA;
aleow@psych.uic.edu}
}




 


\maketitle

\begin{abstract}

Mood disorders are common and associated with significant morbidity and mortality. Early diagnosis has the potential to greatly alleviate the burden of mental illness and the ever increasing costs to families and society. Mobile devices provide us a promising opportunity to detect the users' mood in an unobtrusive manner. In this study, we use a custom keyboard which collects keystrokes' meta-data and accelerometer values. Based on the collected time series data in multiple modalities, we propose a deep personalized mood prediction approach, called {\pro}, by integrating convolutional and recurrent deep architectures  as well as exploring each individual's circadian rhythm. Experimental results not only demonstrate the feasibility and effectiveness of using smart-phone meta-data to predict the presence and severity of mood disturbances in bipolar subjects, but also show the potential of personalized medical treatment for mood disorders. 

\end{abstract}

\begin{IEEEkeywords}
typing dynamics; bipolar disorder; mood prediction; deep learning
\end{IEEEkeywords}

\section{Introduction}

In recent years, people have become more aware of the burden of mental illness and the ever increasing costs to families and society. Mental illness is estimated to be associated with 32.4\% of years lived with disability and 13·0\% of disability-adjusted life-years around the world \cite{vigo2016estimating}, as well as a loss of at least \$193.2 billion in personal earnings in the total
United States population per year \cite{kessler2008individual}. In order to relieve such a crucial situation, it is imperative that we deepen our understanding of mental illness and consequently its diagnosis and treatment. 

Mobile devices have become near ubiquitous with 2 billion smart-phone users worldwide. Smart-phones are used for a variety of tasks including voice/video calling, web browsing, and game playing; however, their most widely-used feature is text messaging \cite{smith2015us}. The frequent utilization of smart-phones in daily lives allows us to study the manifestations of mental illness in an unobtrusive manner.

There have been several studies investigating the feasibility of utilizing smart-phone data as a means to diagnose mood states. Early studies have focused on collecting subjective self-reports of mood from patients \cite{courvoisier2010psychometric,kuntsche2013using,suhara2017deepmood}, which however are subject to biases and may lead to spurious results \cite{bauhoff2011systematic}. More recent studies have focused on the validation of passive data collection methods and yielded promising results in demonstrating the practicality of such methods \cite{faurholt2014smartphone,saeb2015mobile,asselbergs2016mobile,cao2017deepmood}.

Bipolar disorder is a chronic mental illness characterized by recurrent episodes of mania, hypomania, mixed states, and depression \cite{sajatovic2005bipolar} during which changes in neuro-cognitive function, social behavior and circadian activity cycles are well-documented; yet they have not been previously studied using mobile technology ``in the wild'' in the context of keyboard dynamics. We choose to focus on keyboard typing dynamics because features collected from typing are among the most commonly used features in smart-phones, and that by not collecting the actual content of typing the privacy of users is well protected. We hypothesize that depression and mania could be characterized by unique patterns of smart-phone typing dynamics that could be used to predict the severity of these mood states. However, there are several formidable challenges for investigating the relationship between smart-phone typing dynamics and mood states:
\begin{itemize}
\item\textbf{Local patterns}: Typing data is similar to text sequences in that they both have similar local patterns, such as n-gram phase that represents a certain meaningful concept. Besides alphanumeric data, the accelerometer data may also have local patterns that reveal the situation under which the keys are pressed. How to capture such local patterns is thus a problem that must be handled.
\item\textbf{Circadian rhythm}: Circadian rhythm is a biological process that displays an endogenous oscillation of about 24 hours \cite{edgar2012peroxiredoxins}\cite{althoff2017harnessing}. It is running in the background of human brain and cycles between sleepiness and alertness at regular intervals. It is commonly known that individual depressive moods vary according to the circadian rhythm, as well as the day of the week \cite{suhara2017deepmood}. Thus it is important to consider the effects of circadian rhythm.
\item\textbf{Personalization}: Different smart-phone users usually have very different typing behaviors. Moreover, each person may have different baselines in terms of mood states, even with similar typing dynamics. Therefore, we should make personalized mood prediction rather than using a subject-unaware model.
\end{itemize}

In this paper, we propose a deep architecture based on late fusion, named {\pro}, to model smart-phone typing dynamics, as illustrated in Figure~\ref{fig:architecture}. The contributions of this work are as follow:
\begin{itemize}
\item\textbf{Early fusion of features}: When dealing with multi-view data, features from different views are usually fused at a late stage of the model~\cite{cao2017deepmood}, but we explore two ways of performing early fusion and find that they both outperform the late fusion approach.
\item\textbf{Stacking CNNs and RNNs for mood prediction}: In conjunction with recurrent neural networks that captures overall temporal dynamics, convolutional neural networks are used to capture the local patterns in the typing dynamics.
\item\textbf{Time-based calibration per person}: The final prediction is calibrated per person using the temporal information in order to exploit the circadian rhythm and achieve personalized mood prediction, or in general, \emph{precision medicine}. This enables us to improve prevention strategies and treatments by better incorporating individual patient characteristics.
\end{itemize}

We conduct experiments showing that our model achieves the lowest regression error on Hamilton Depression Rating Scale (HDRS) \cite{williams1988structured} and Young Mania Rating Scale (YMRS) \cite{young1978rating} prediction tasks, which indicates the potential of using smart-phone meta-data to predict mood disturbance and severity. We also study the effect of early fusion in mood prediction and find that early fusion works better than late fusion approach. Our code is open-source at \textbf{\url{https://github.com/stevehuanghe/dpMood}}.

\section{Data Analysis}
\label{sec:data}

The data used in this work is obtained from the BiAffect\footnote{\url{http://www.biaffect.com/}} project by the team's permission for research use. The data was collected in the following way. During a preliminary data collection phase, for a period of 8 weeks, 40 individuals were provided a Galaxy Note 4 mobile phone which they were instructed to use as their primary phone during the study. This phone was loaded with a custom keyboard that replaced the standard Android OS keyboard. The keyboard collected meta-data consisting of keypress entry time and accelerometer movement and uploaded them to the web server. In order to protect subjects' privacy, the detailed content of typing with the exceptions of the backspace key and space bar was not collected.

\begin{table}[t]
\caption{Demographic characteristics.}
\small
\label{tab:demo}
\centering
\begin{tabular}{lccc}
\toprule
& Bipolar I & Bipolar II & Control\\
\midrule 
Age (years) & 45.6 $\pm$ 9.9 & 52.4 $\pm$ 9.4 & 46.1 $\pm$ 10.7 \\
Gender (\% female) & 57\% & 80\% & 63\% \\
Years of education & 15.4 $\pm$ 1.7 & 14.8 $\pm$ 2.8 & 15.8 $\pm$ 1.4 \\
IQ & 109.0 $\pm$ 3.4 & 102.2 $\pm$ 8.5 & 107.8 $\pm$ 11.7 \\
\bottomrule
\end{tabular}
\end{table}

As in \cite{cao2017deepmood}, we study the collected meta-data of bipolar subjects and normal controls who had provided at least one week of meta-data. In the selected subset of data, there are 7 subjects with {\em bipolar I} disorder that involves periods of severe mood episodes from mania to depression, 5 subjects with {\em bipolar II} disorder which is a milder form of mood elevation, involving milder episodes of hypomania that alternate with periods of severe depression, and 8 subjects with no diagnosis per DSM-IV TR criteria \cite{kessler2005lifetime}. Subjects were administered the Hamilton Depression Rating Scale (HDRS) \cite{williams1988structured} and Young Mania Rating Scale (YMRS) \cite{young1978rating} once a week which are used as the golden standard to assess the level of depressive and manic symptoms in bipolar disorder. Also, subjects performed self-evaluations everyday. Subject demographics are described in Table~\ref{tab:demo}.

Two groups of features are used in this work: alphanumeric characters and accelerometer values. Although special characters were investigated in \cite{cao2017deepmood}, we find that their contribution in our model performance is negligible and will even cause a decrease on the depression prediction task by about 2\%. For alphanumeric characters, due to privacy reasons, we only collected meta-data for these keypresses, including duration of a keypress, time since last keypress, and distance from last key along two axises. On the other hand, accelerometer values along three axises are recorded every 60ms in the background during an active session regardless of a person's typing speed.

\begin{figure}[t]
\centering
\begin{minipage}[l]{0.9\columnwidth}
\centering
\includegraphics[width=1\textwidth]{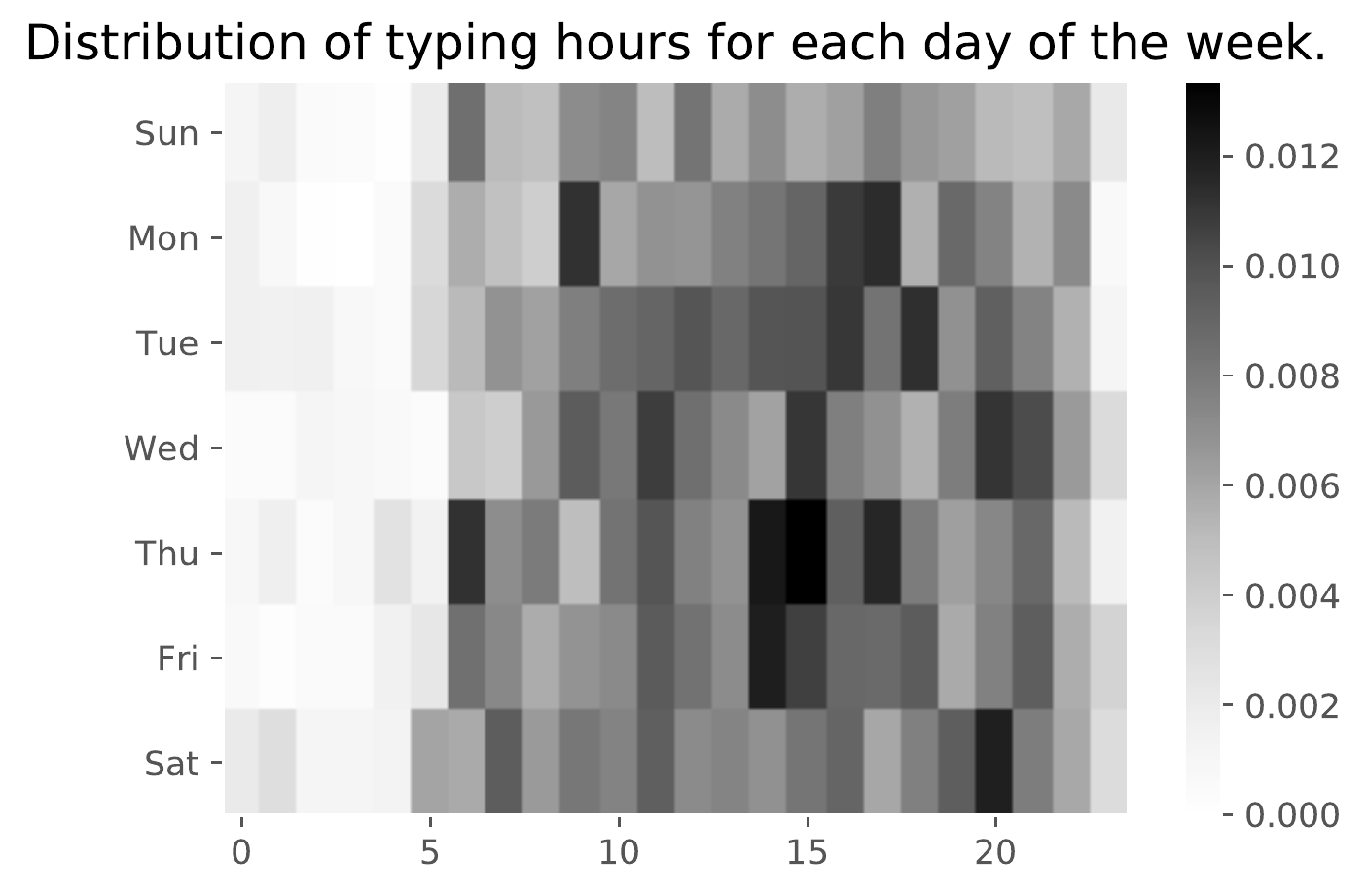}
\end{minipage}
\caption{Distribution of typing hours for each day of the week.}
\label{fig:dist_day_hour}
\end{figure}

\begin{figure}[t]
\centering
\subfigure[]{
\begin{minipage}[l]{0.45\columnwidth}
\centering
\includegraphics[width=0.9\textwidth]{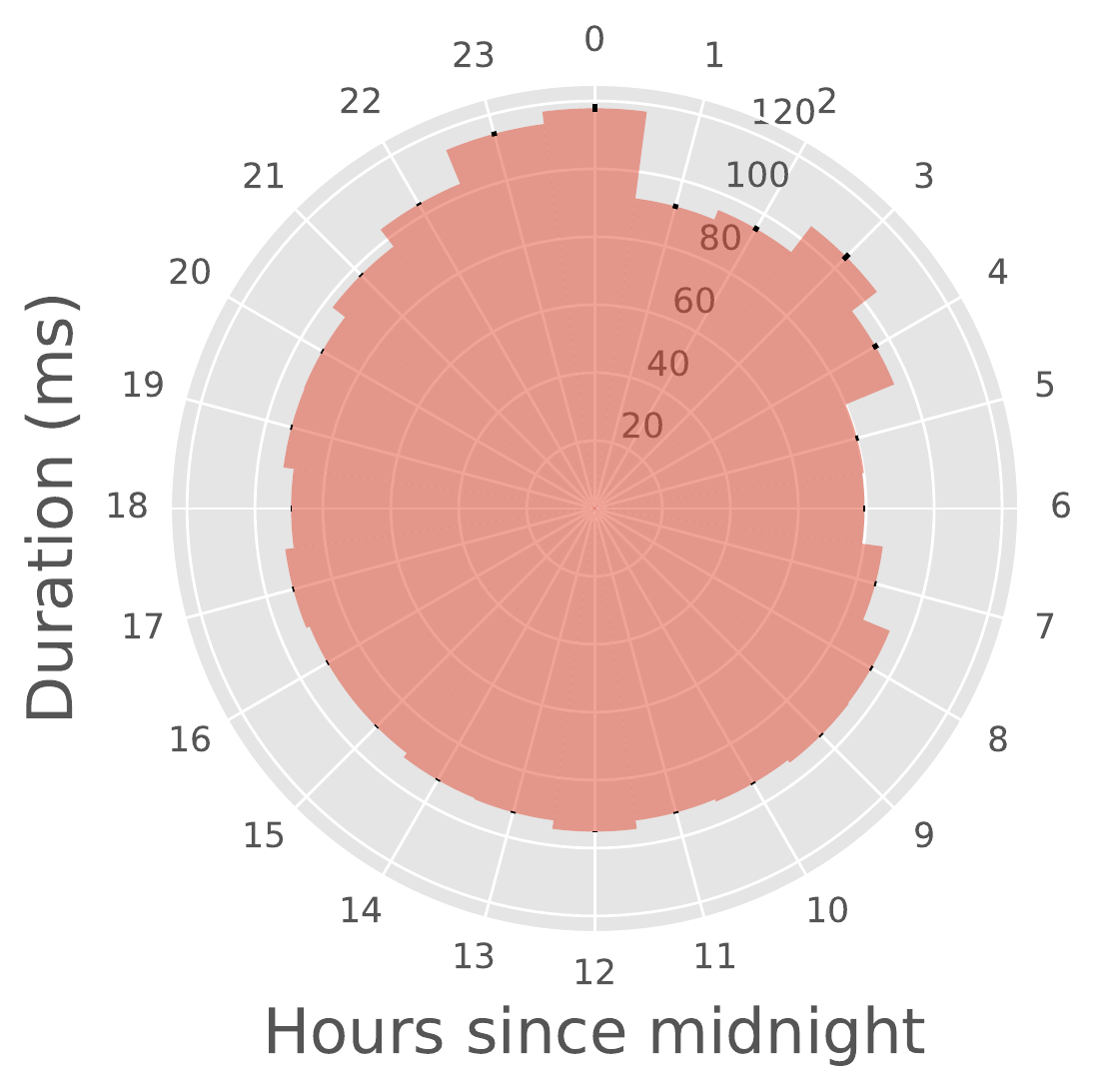}
\end{minipage}
\label{fig:circadian_duration}
}
\subfigure[]{
\begin{minipage}[l]{0.45\columnwidth}
\centering
\includegraphics[width=0.82\textwidth]{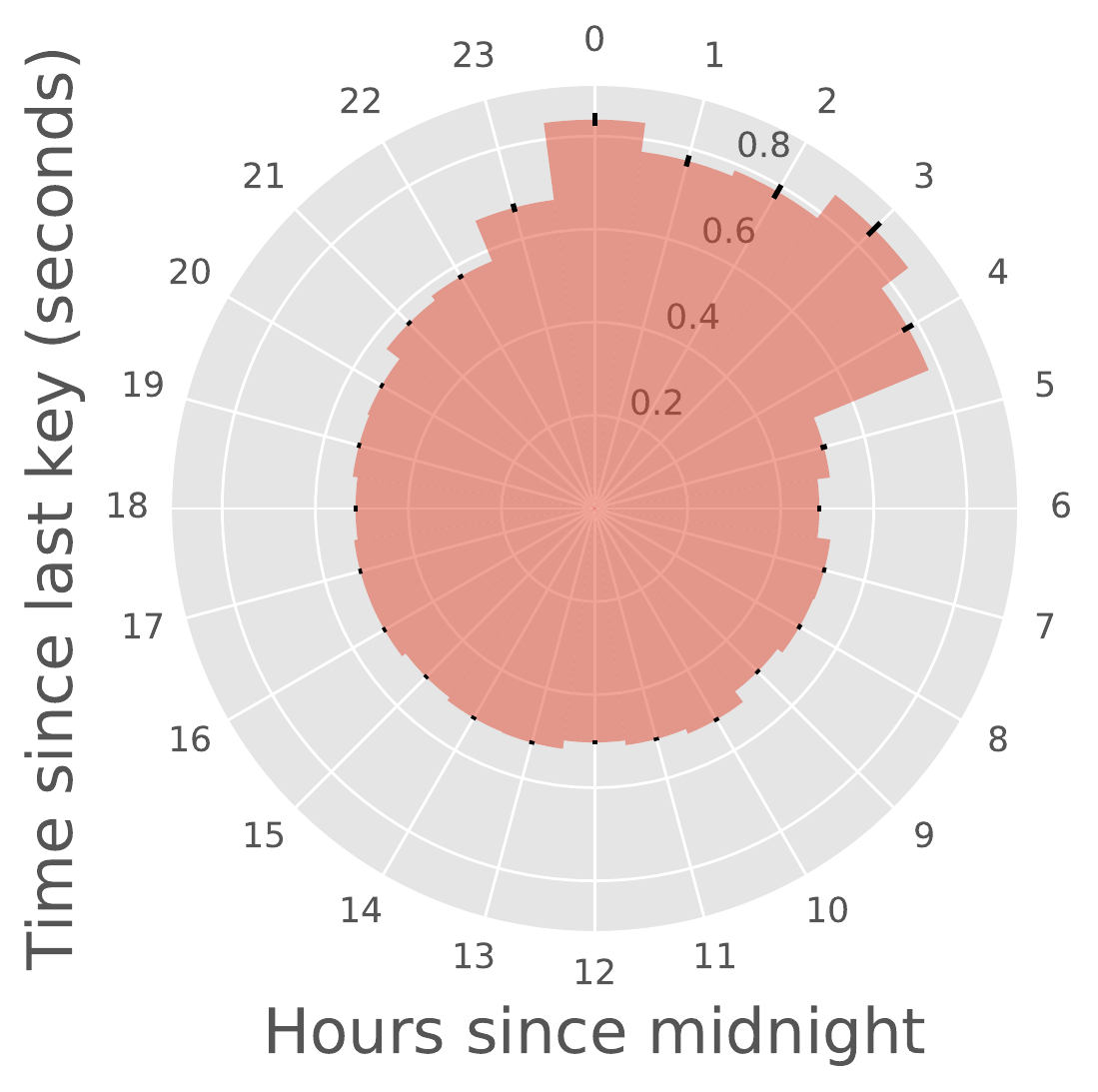}
\end{minipage}
\label{fig:circadian_timesincelastkey}
}
\subfigure[]{
\begin{minipage}[l]{0.9\columnwidth}
\centering
\includegraphics[width=1\textwidth]{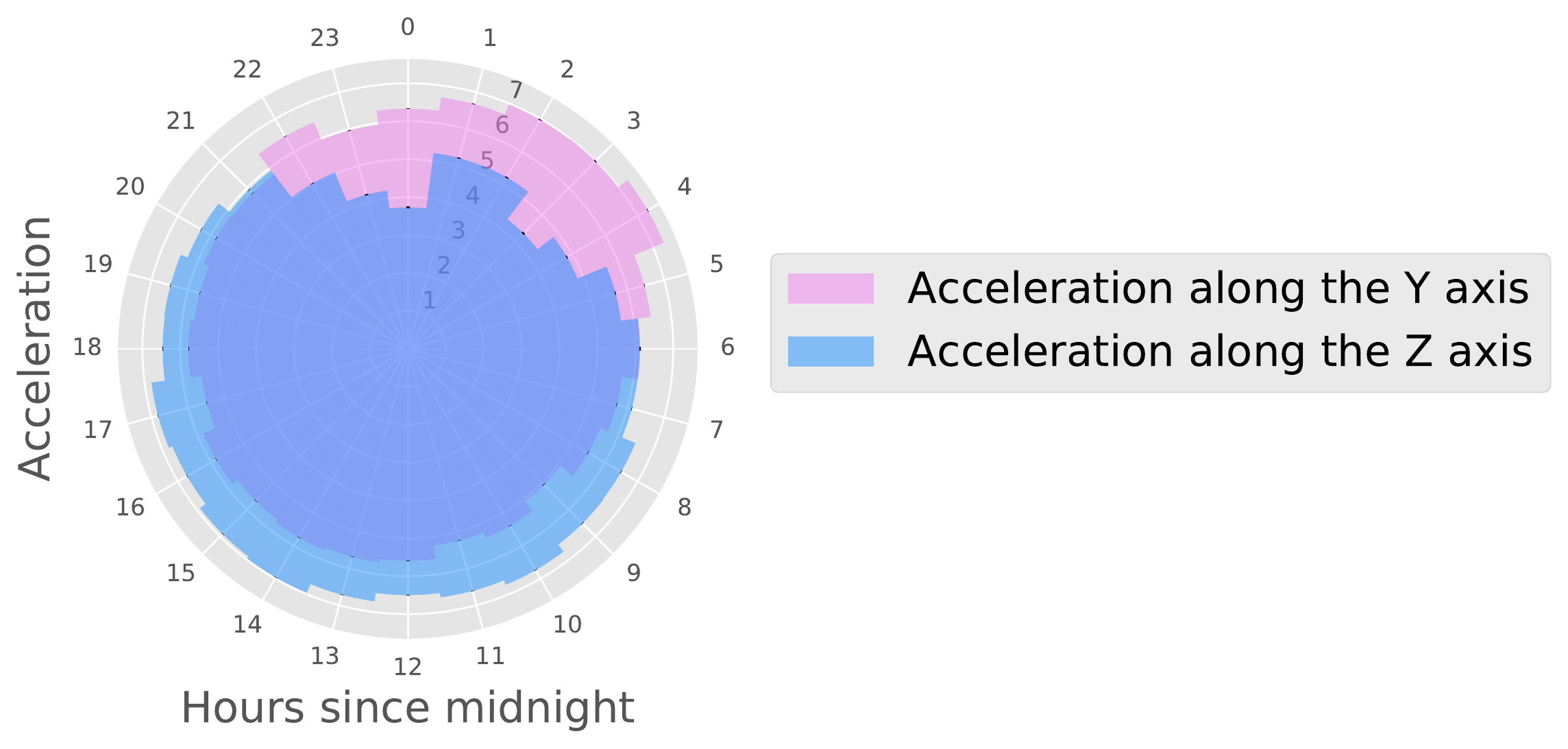}
\end{minipage}
\label{fig:circadian_accel}
}
\caption{Circadian rhythm.}
\label{fig:circadian}
\end{figure}

\begin{figure}[t]
\centering
\subfigure[]{
\begin{minipage}[l]{0.45\columnwidth}
\centering
\includegraphics[width=0.9\textwidth]{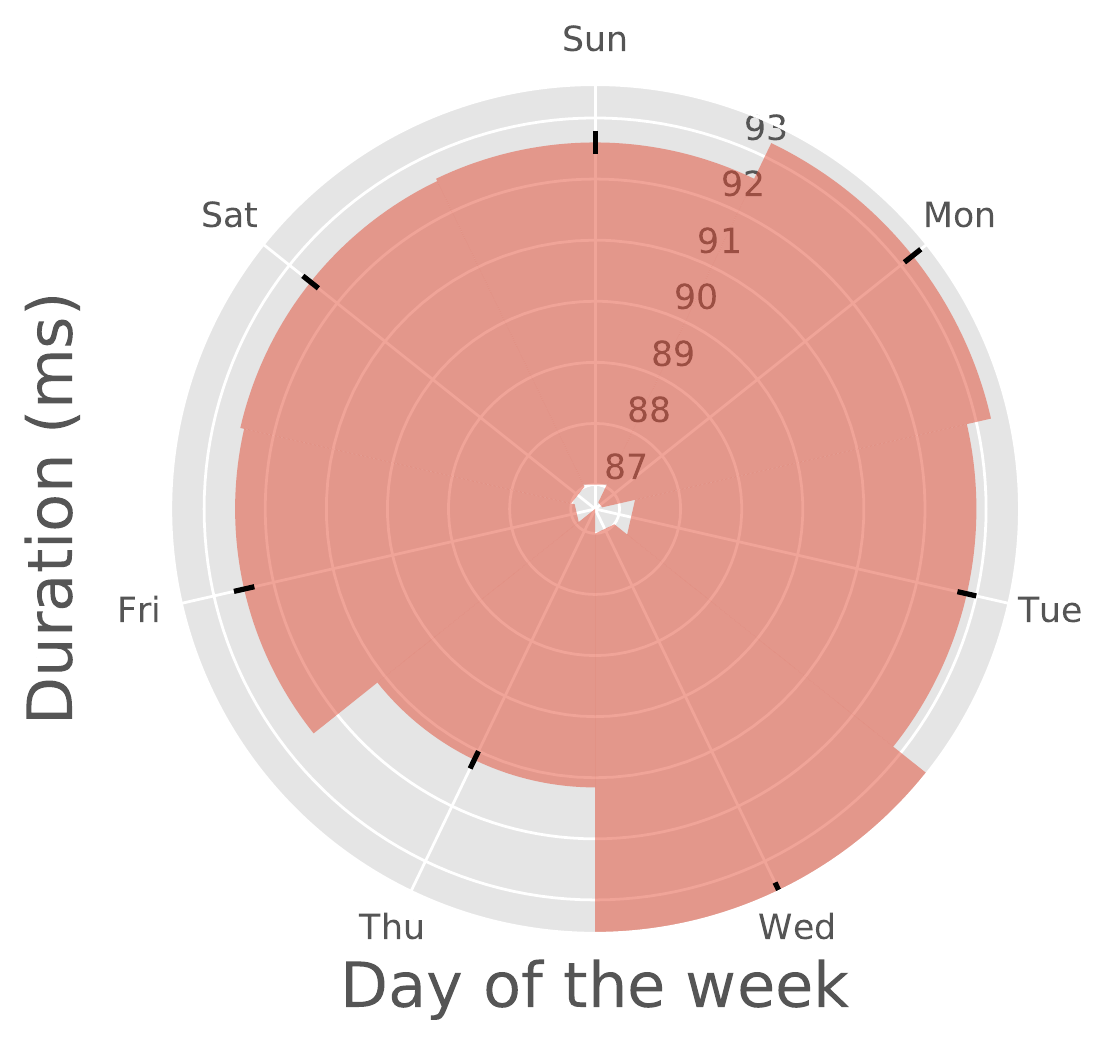}
\end{minipage}
\label{fig:dayofweek_duration}
}
\subfigure[]{
\begin{minipage}[l]{0.45\columnwidth}
\centering
\includegraphics[width=0.9\textwidth]{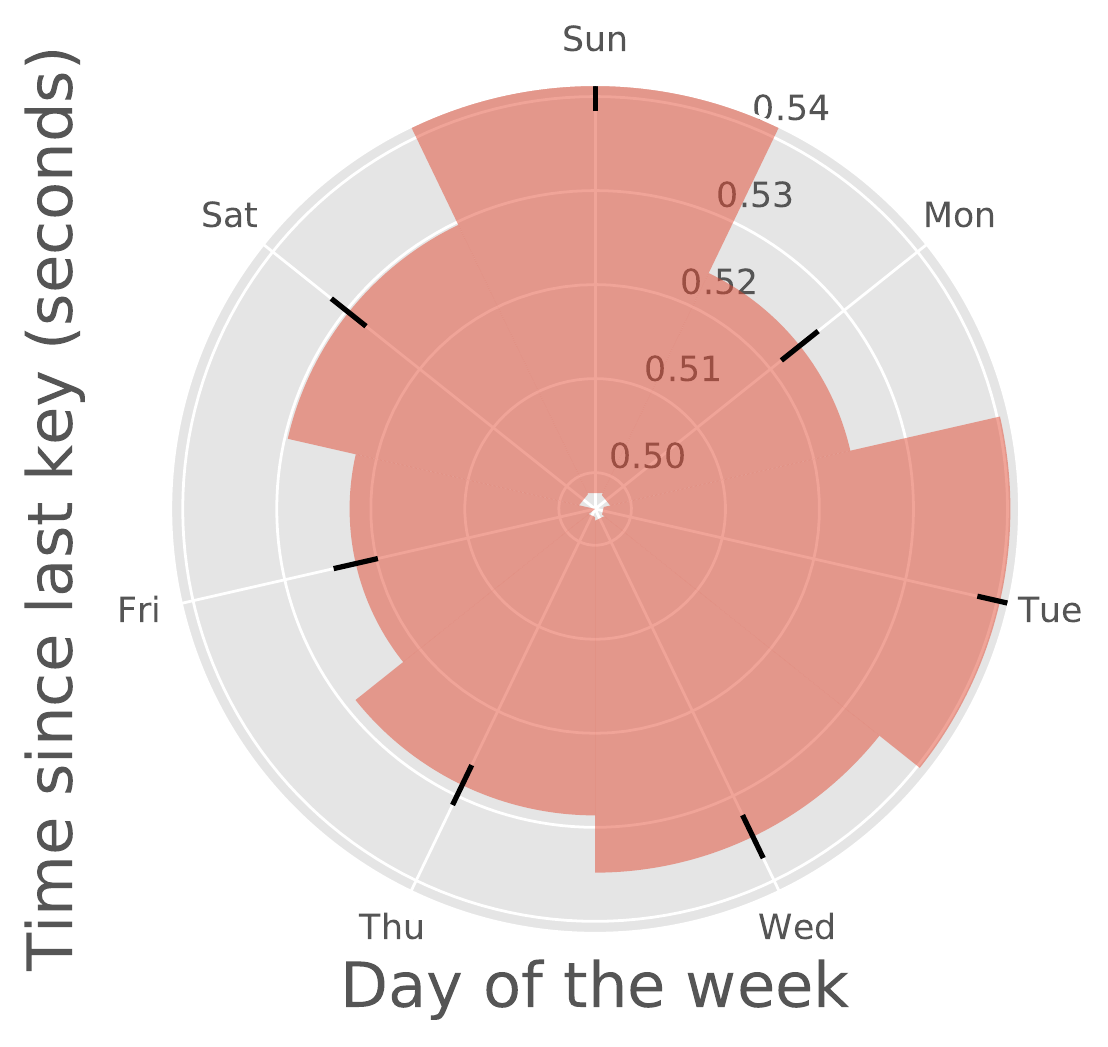}
\end{minipage}
\label{fig:dayofweek_timesincelastkey}
}
\subfigure[]{
\begin{minipage}[l]{0.45\columnwidth}
\centering
\includegraphics[width=0.9\textwidth]{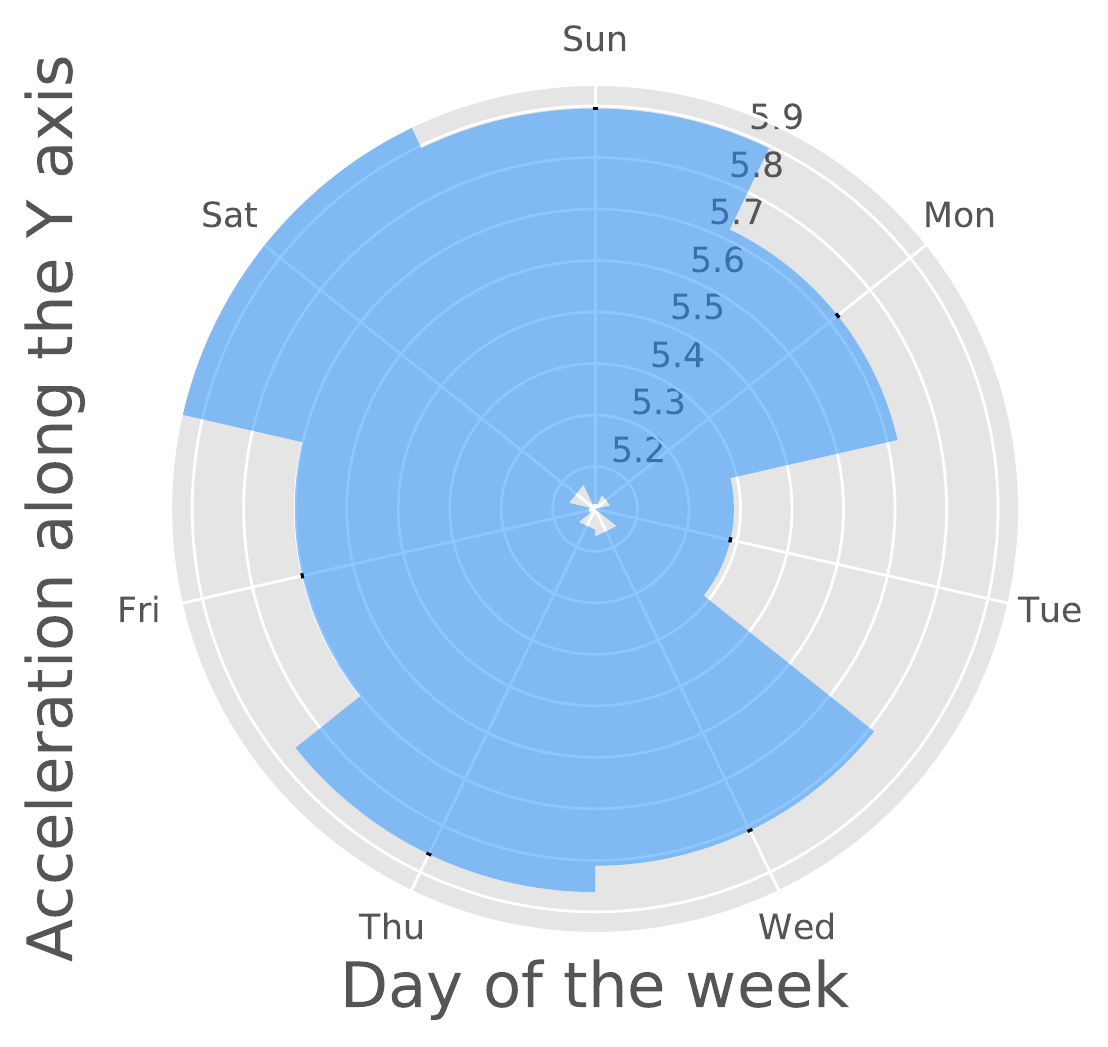}
\end{minipage}
\label{fig:dayofweek_accely}
}
\subfigure[]{
\begin{minipage}[l]{0.45\columnwidth}
\centering
\includegraphics[width=0.9\textwidth]{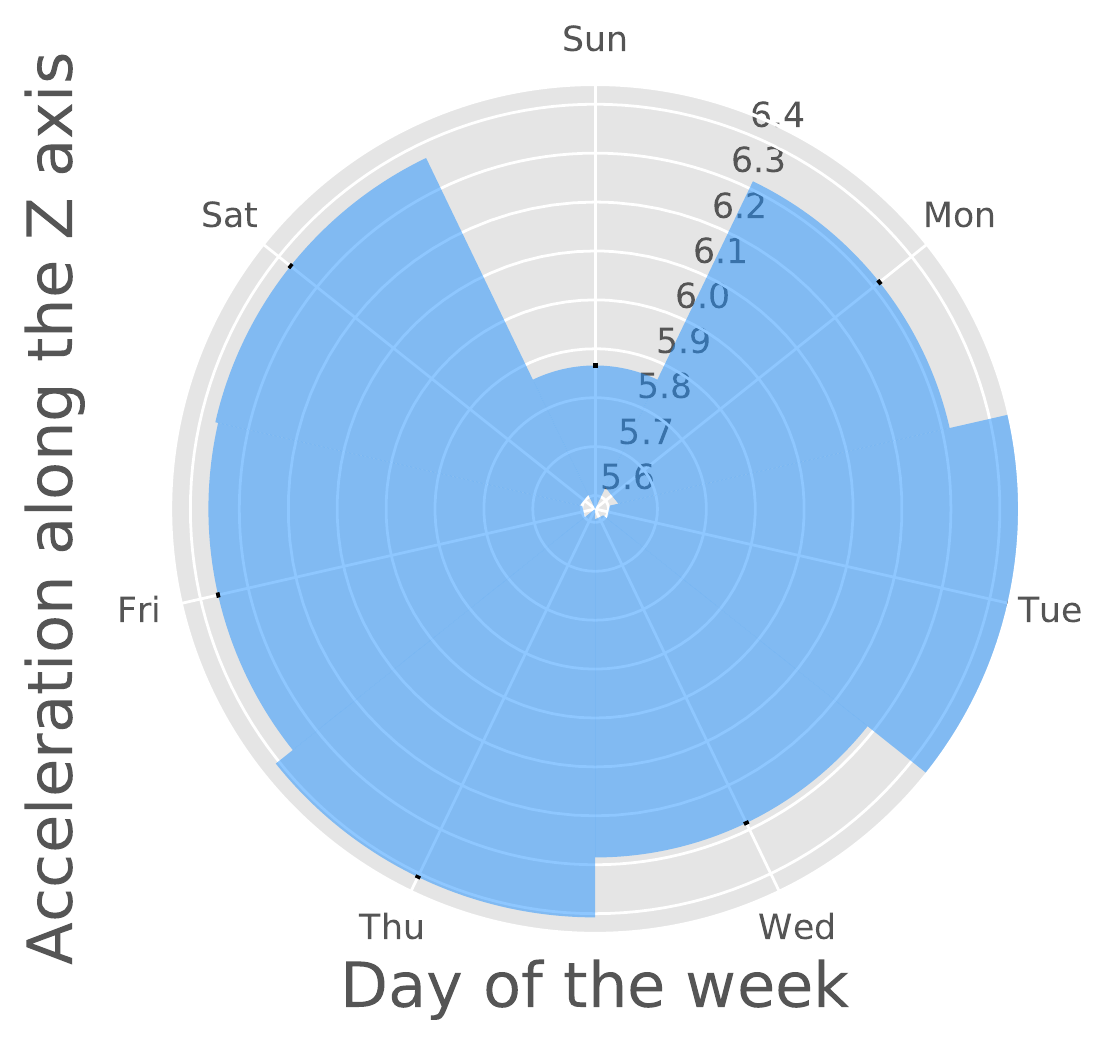}
\end{minipage}
\label{fig:dayofweek_accelz}
}
\caption{Day of the week.}
\label{fig:dayofweek}
\end{figure}

\subsection{Time Effects}
\label{sec:data_time}

Although Cao et al.~\cite{cao2017deepmood} provide data analysis on the correlation between patterns of typing meta-data and mood in bipolar disorder, they do not study the temporal effects on typing dynamics. Here we investigate the relationship between the typing dynamics and time, in order to justify the necessity of time-based calibration in our model. Figure~\ref{fig:dist_day_hour} presents the distribution of typing hours in the 7-day by 24-hour matrix.

Figure~\ref{fig:circadian} shows how the mean and standard deviation values of some features change over 24 hours. They are computed from all the samples in the dataset. About alphanumeric characters, we can observe that the {\em duration of a keypress} (Figure~\ref{fig:circadian_duration}) and the {\em time since last keypress} (Figure~\ref{fig:circadian_timesincelastkey}) are correlated and share the same pattern. In general, the fastest typing speed occurs at \texttt{6:00} and remains stable between \texttt{8:00} to \texttt{20:00}, and then it becomes significantly slower during the midnight. We suspect that this is primarily due to the circadian rhythm which is a biological process that displays an endogenous oscillation of about 24 hours \cite{edgar2012peroxiredoxins}. It is running in the background of human brain and cycles between sleepiness and alertness at regular intervals.

About accelerometer values, we omit the acceleration along the X axis here, because data were collected only when the phone was in a portrait position which makes the X-axis acceleration to be centered around 0 and less interesting. Figure~\ref{fig:circadian_accel} reveals a complementary relationship between the acceleration along the Y axis and that along the Z axis. The fact that the Z-axis acceleration will usually be negative when one is lying and using the phone may explain the observation that the average Z-axis acceleration is small (large Y-axis acceleration) during the night time, and there is relatively large Z-axis acceleration (small Y-axis acceleration) during the day time.

Previous studies have shown that people's mental health depends on the day of the week \cite{suhara2017deepmood,golder2011diurnal}. In Figure~\ref{fig:dayofweek}, we illustrate how these typing dynamics features correlate with the day of the week. It can be seen that the duration of a keypress and the time since last keypress are significantly different across different days during a week, although they are not as interpretable as with the circadian rhythm. Moreover, accelerometer values along Y and Z axises also vary on different days of the week, and we suspect that the relatively smaller Z-axis acceleration and larger Y-axis acceleration on Sunday may result from that people spend more time lying in the bed or couch on Sunday. It motivates us to incorporate the time effects into the model design.


\begin{figure*}[t]
\centering
\subfigure[]{
\begin{minipage}[l]{0.45\columnwidth}
\centering
\includegraphics[width=1\textwidth]{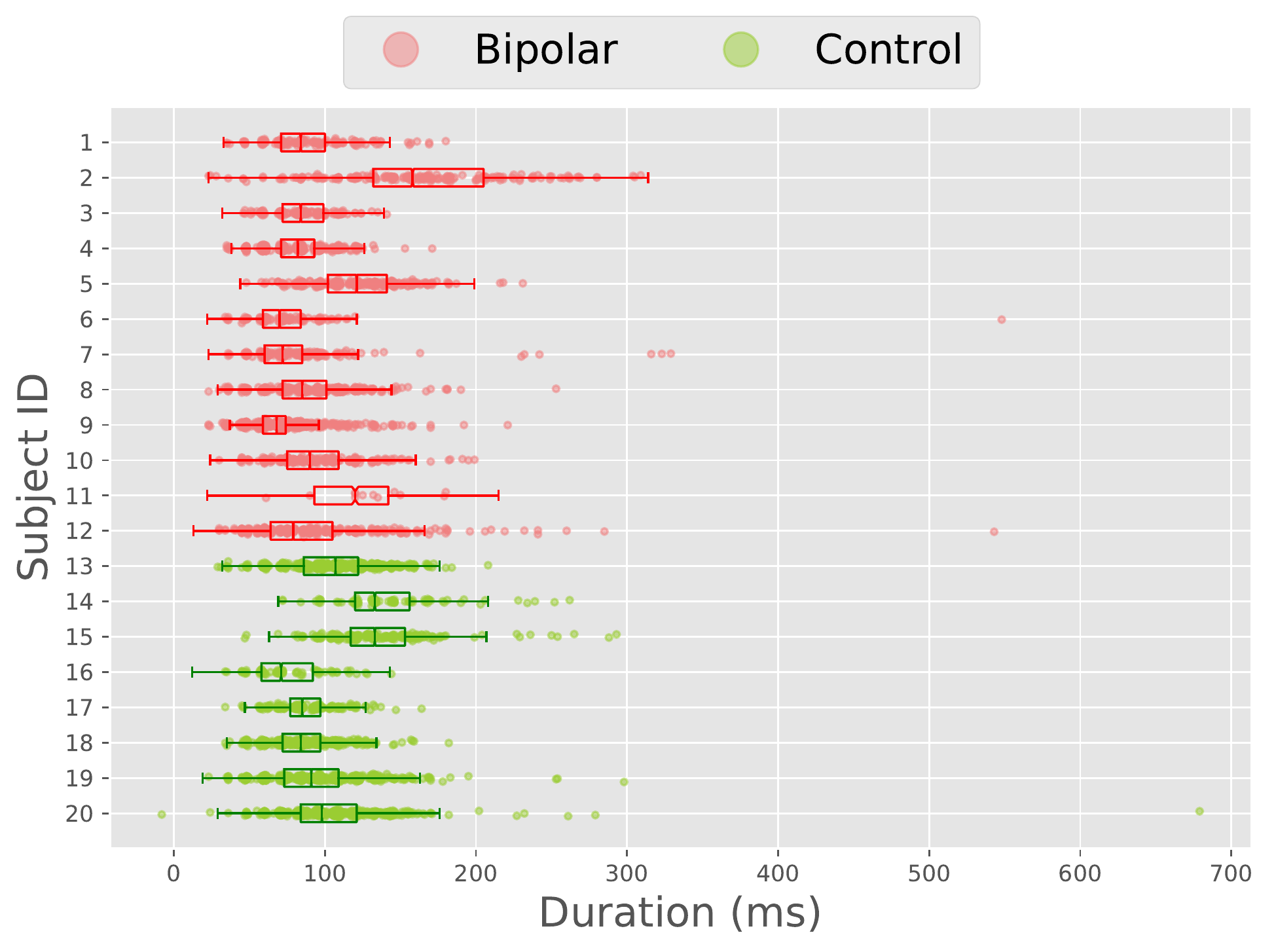}
\end{minipage}
\label{fig:subject_duration}
}
\subfigure[]{
\begin{minipage}[l]{0.45\columnwidth}
\centering
\includegraphics[width=1\textwidth]{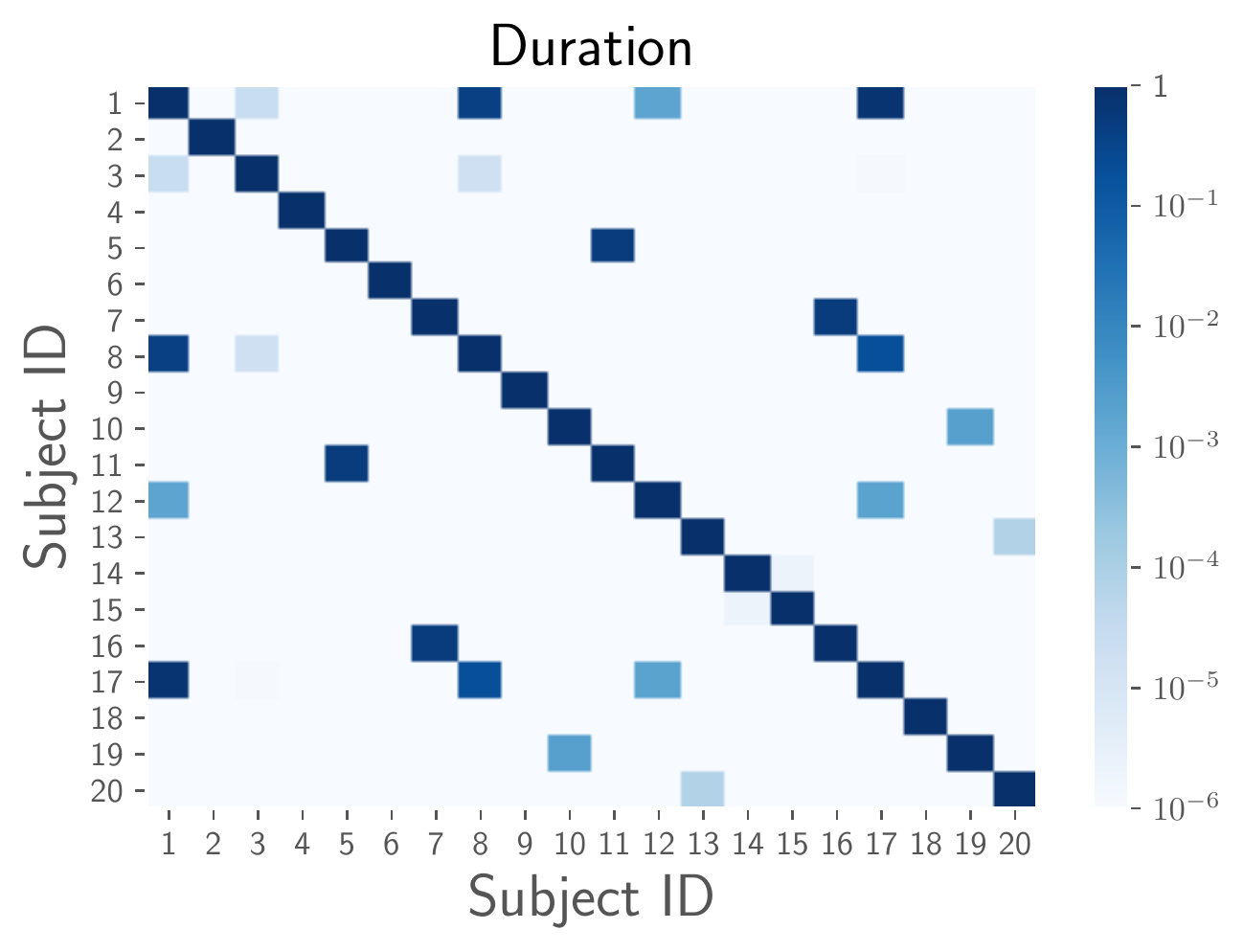}
\end{minipage}
\label{fig:ttest_duration}
}
\subfigure[]{
\begin{minipage}[l]{0.45\columnwidth}
\centering
\includegraphics[width=1\textwidth]{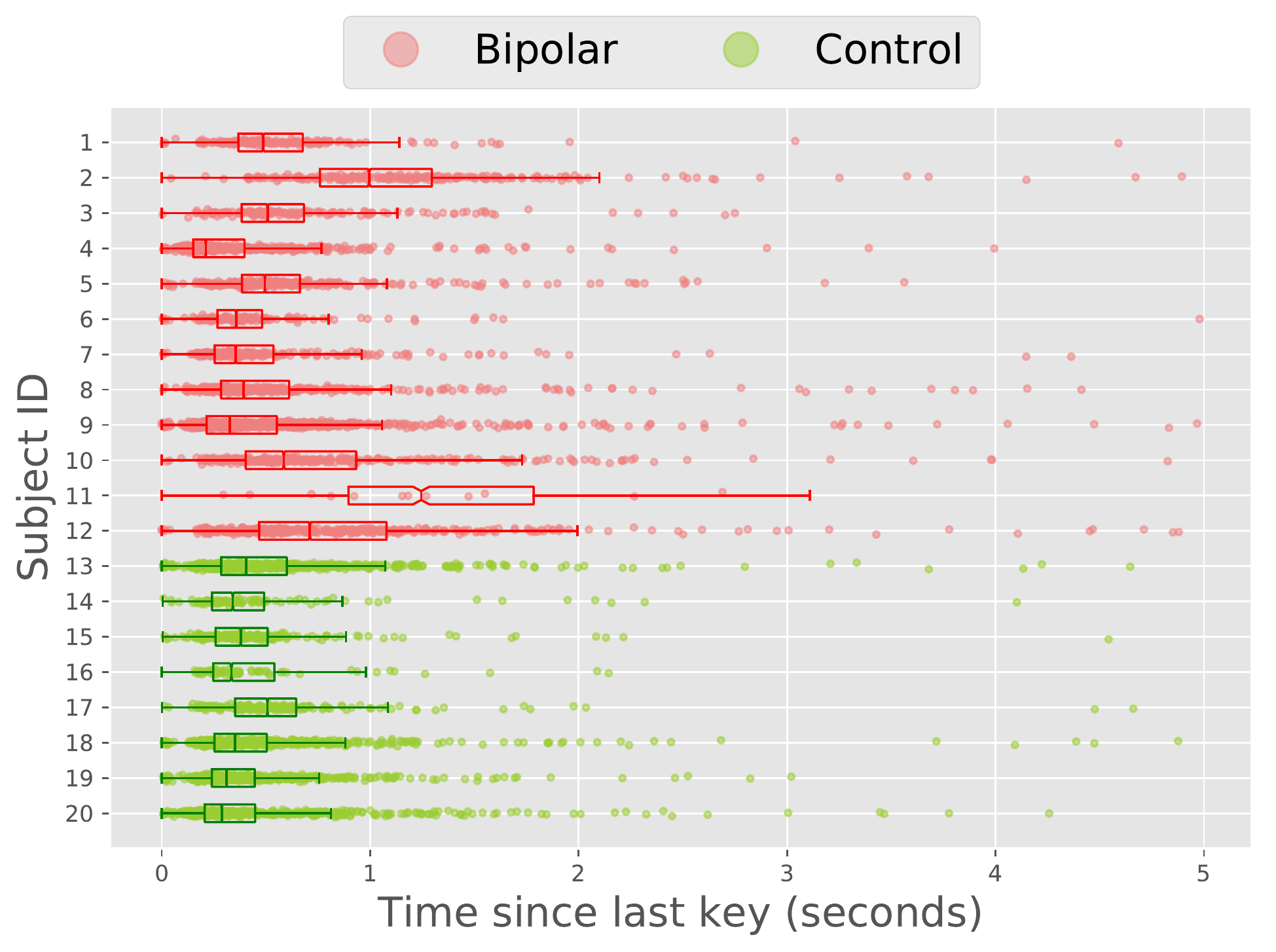}
\end{minipage}
\label{fig:subject_timesincelastkey}
}
\subfigure[]{
\begin{minipage}[l]{0.45\columnwidth}
\centering
\includegraphics[width=1\textwidth]{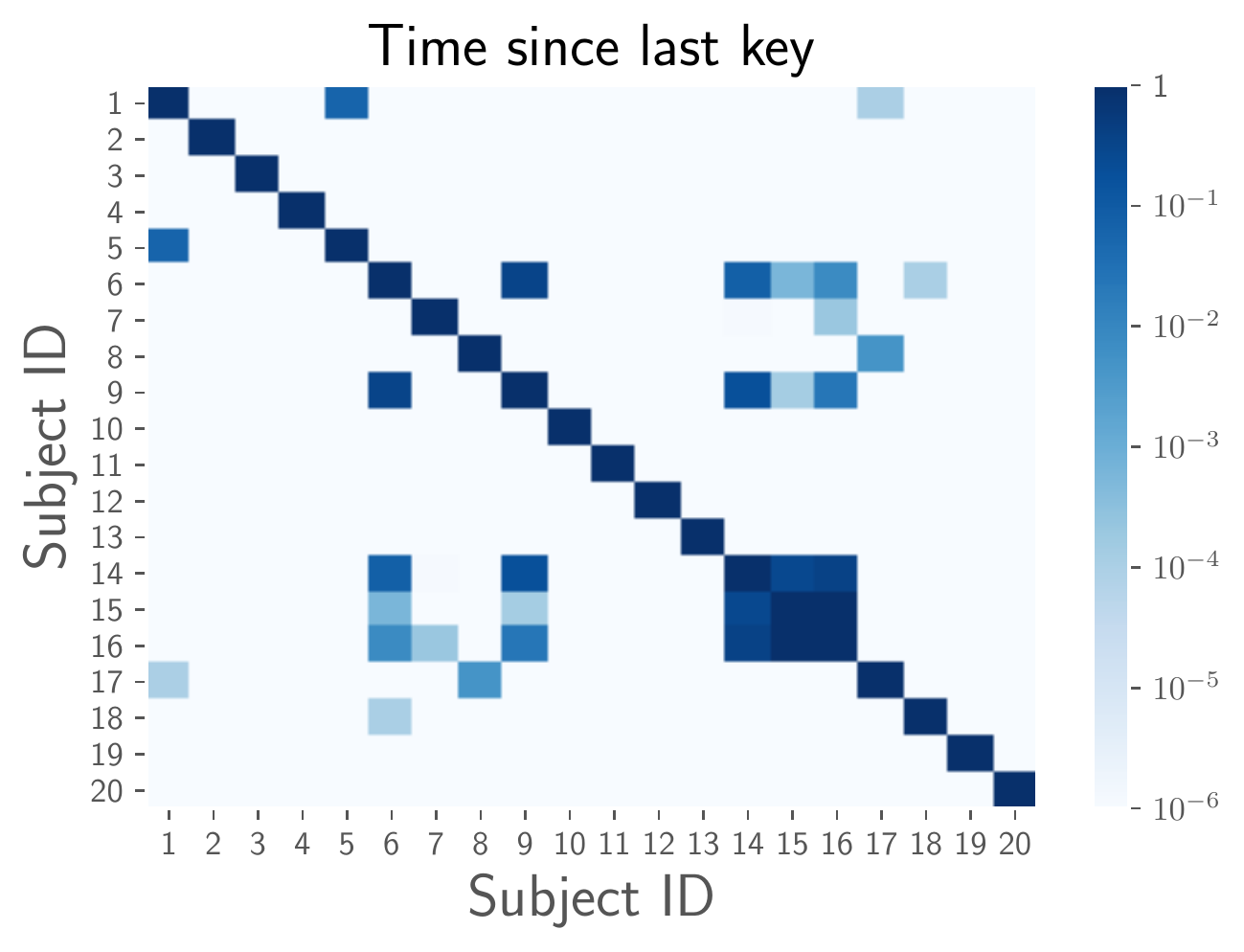}
\end{minipage}
\label{fig:ttest_timesincelastkey}
}
\subfigure[]{
\begin{minipage}[l]{0.45\columnwidth}
\centering
\includegraphics[width=1\textwidth]{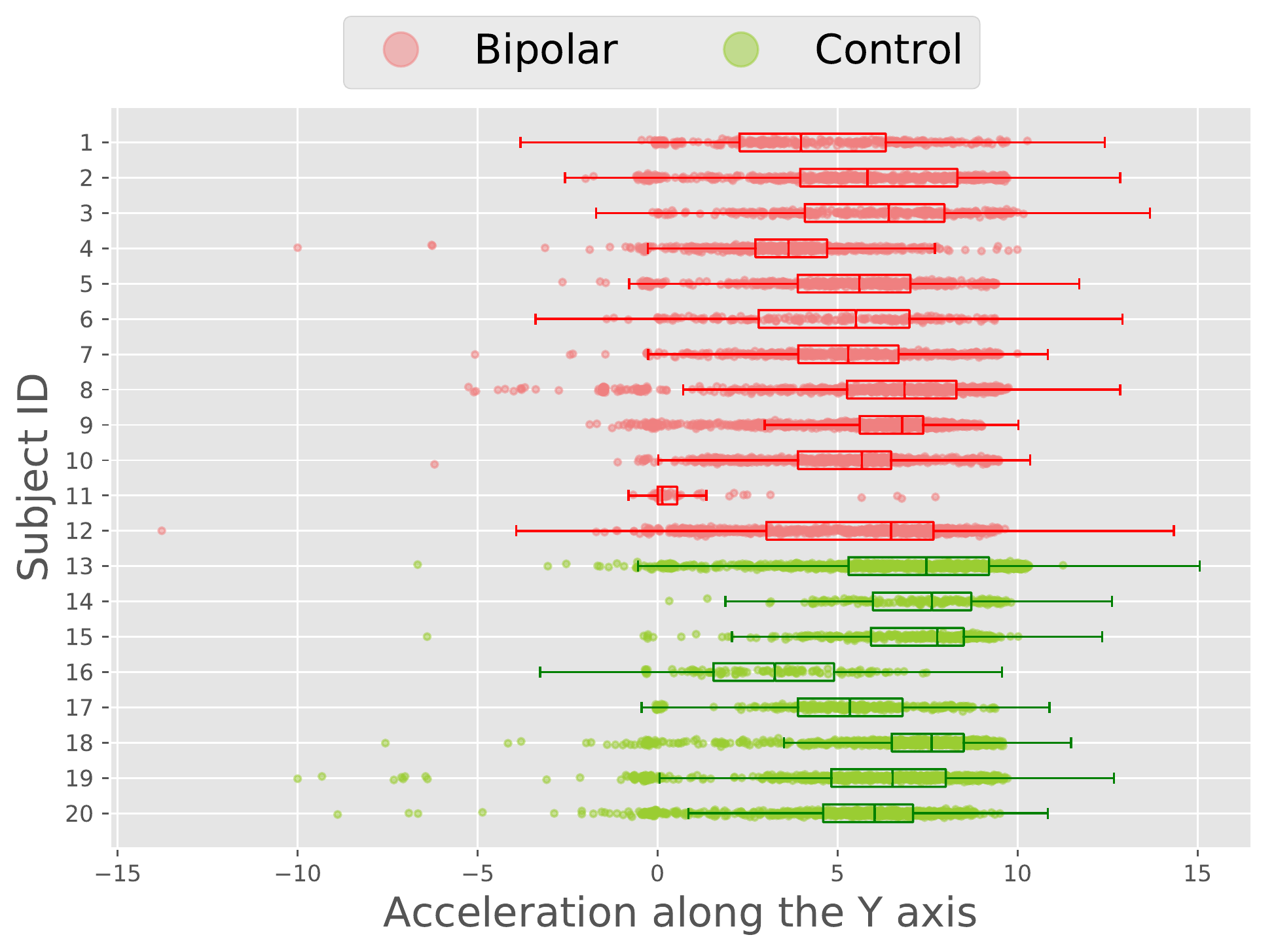}
\end{minipage}
\label{fig:subject_accely}
}
\subfigure[]{
\begin{minipage}[l]{0.45\columnwidth}
\centering
\includegraphics[width=1\textwidth]{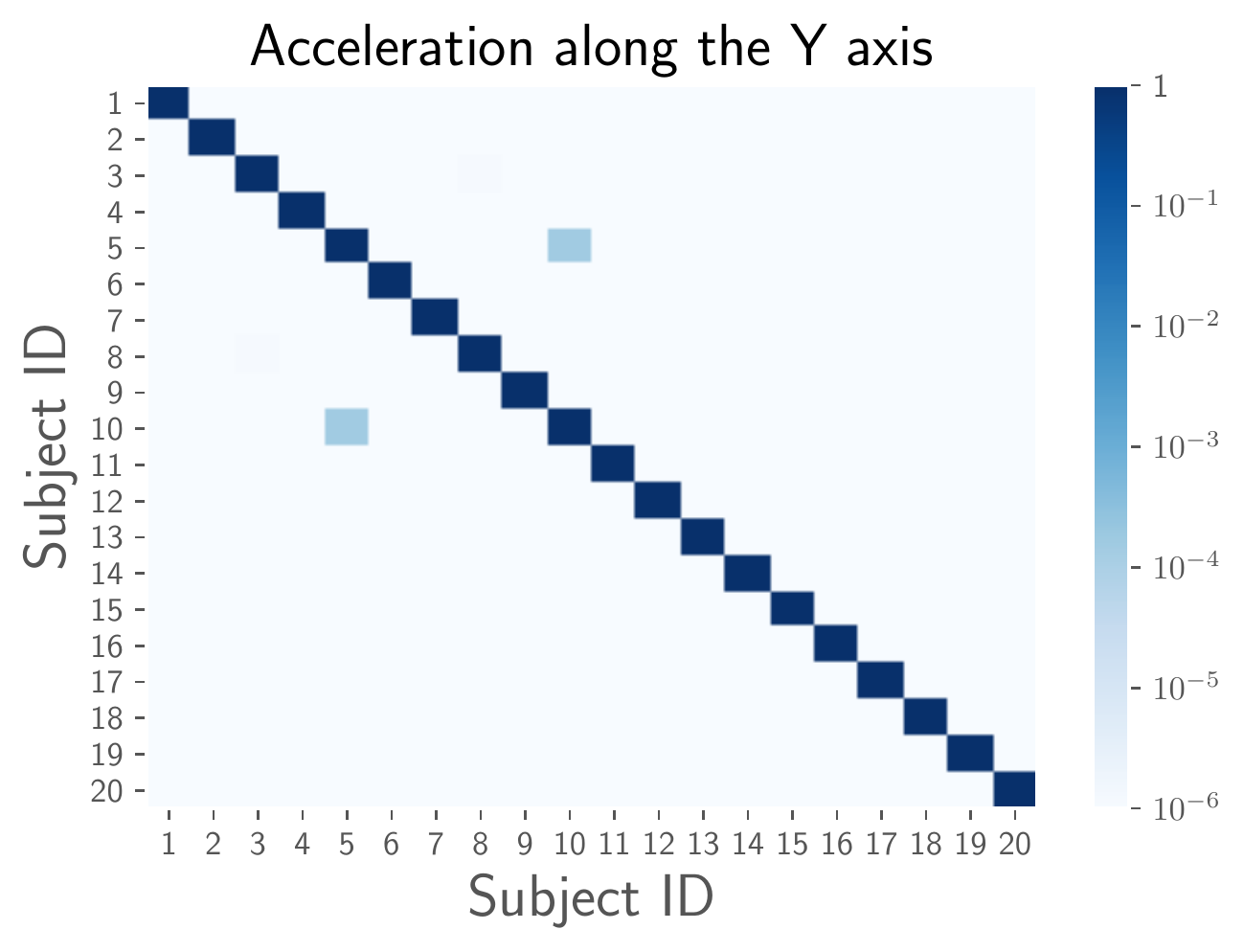}
\end{minipage}
\label{fig:ttest_accely}
}
\subfigure[]{
\begin{minipage}[l]{0.45\columnwidth}
\centering
\includegraphics[width=1\textwidth]{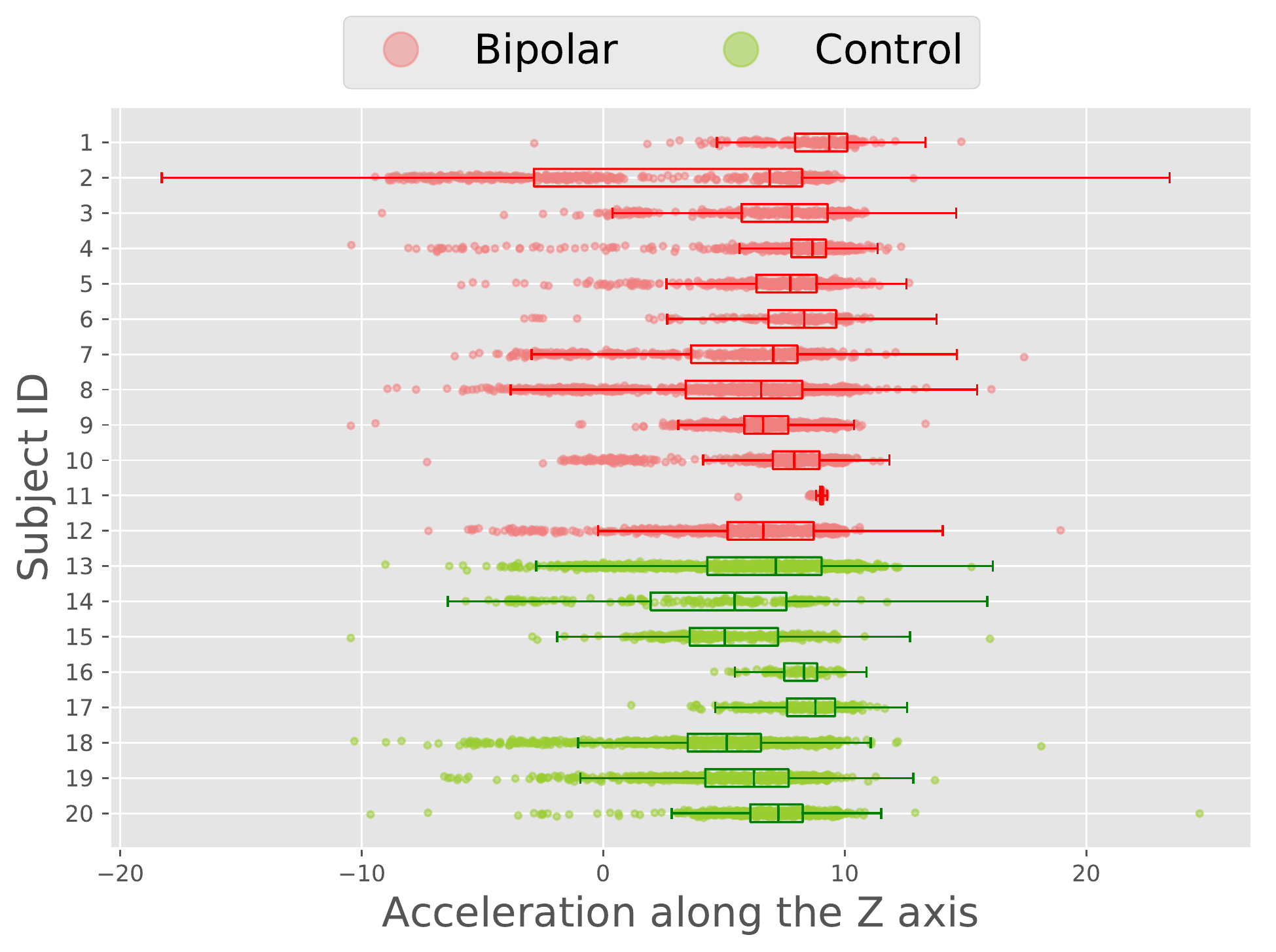}
\end{minipage}
\label{fig:subject_accelz}
}
\subfigure[]{
\begin{minipage}[l]{0.45\columnwidth}
\centering
\includegraphics[width=1\textwidth]{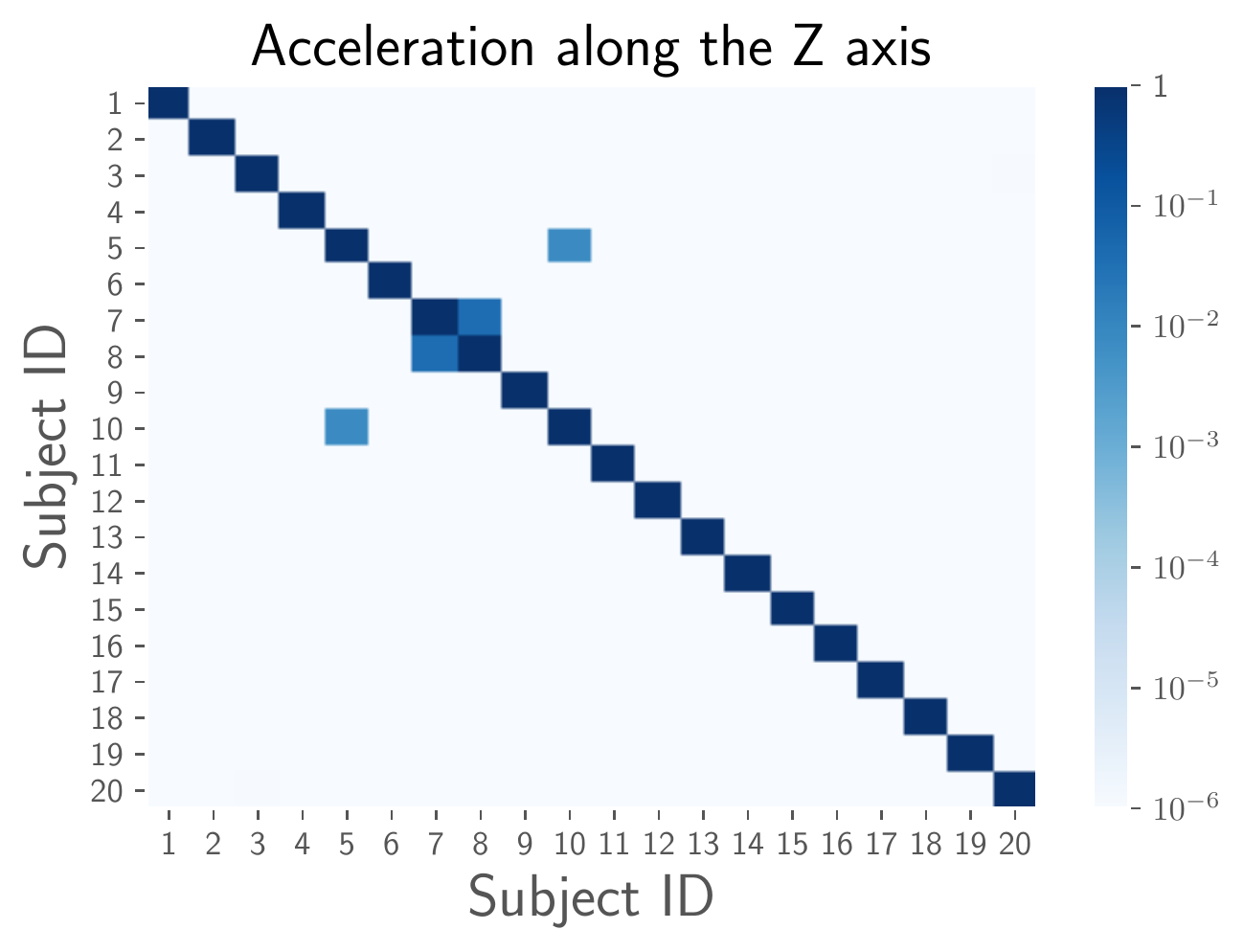}
\end{minipage}
\label{fig:ttest_accelz}
}
\caption{Personal traits. Left column: boxplots with scattered samples per subject. Right column: heatmaps of p-values from subject-level two-sided T-tests (a white cell indicates that the corresponding pair of subjects are significantly different in terms of the feature distribution).}
\label{fig:subject}
\end{figure*}

\subsection{Personal Traits}
\label{sec:data_subject}


We further explore the personal traits hidden in the typing dynamics by analyzing the feature distributions across subjects, as shown in Figure~\ref{fig:subject} where the left column (Figure~\ref{fig:subject_duration},~\ref{fig:subject_timesincelastkey},~\ref{fig:subject_accely},~\ref{fig:subject_accelz}) presents the boxplots with scattered samples per subject of {\em duration of a keypress}, {\em time since last keypress}, {\em acceleration along the Y axis}, and {\em acceleration along the Z axis}, and in the right column (Figure~\ref{fig:ttest_duration},~\ref{fig:ttest_timesincelastkey},~\ref{fig:ttest_accely},~\ref{fig:ttest_accelz}), there are the corresponding heatmaps showing the p-values from the subject-level two-sided T-tests. We conclude that two subjects often have significantly different feature distributions regardless of their diagnosis. Therefore, it is critical to consider the subject-level baselines.

Moreover, it is worthwhile noting that although the connection between the bipolar disorder and these typing behavior features is not obvious from Figure~\ref{fig:subject}, we will show in experiments that accurate mood prediction can be achieved by capturing the local patterns, exploiting the circadian rhythm, and personalization in our proposed model.

\section{Preliminaries}
There are two key modules used in this work, the temporal convolutional neural network and the Gated Recurrent Unit (GRU), which are briefly reviewed in this section.

\subsection{Temporal Convolution}

Although convolutional neural networks (CNNs) are most widely used in computer vision models, it has recently been demonstrated that CNNs also work well on some natural language processing (NLP) tasks such as sentence classification \cite{kim2014convolutional,zhang2015character}.

A temporal convolutional module computes an 1-dimensional convolution over an input sequence. Given an input sequence $f(x)=\{f_1(x), f_2(x),..., f_n(x)\}$ with length $l$ and $n$ channels, and a list of $m$ discrete kernels $g(x)=\{g_1(X), g_2(x),...,g_m(x)\}$ with length $k$, then the convolution between $f$ and $g$ with stride $d$ is calculated by:
\begin{align}
    h_j(y) = \sum_{i=1}^n \sum_{x=1}^k{g_j(x) \cdot f_i(y \cdot d - x + c)},
\end{align}
where $h(y)=\{h_1(y), h_2(y),...,h_m(y)\}$, $c=k-d+1$ is an offset constant, and the final output feature $h(y)$ is of size $[(l-k+1)/d, m]$.

\subsection{Gated Recurrent Unit}

Recurrent neural networks (RNNs) are widely used to model the dynamics of sequence data where hidden states are updated to model the changes in the sequence over time.The current hidden state $\textbf{h}_t$ is computed using the current input $\textbf{x}_t$ and the previous hidden state $\textbf{h}_{t-1}$. The simplest way of implementing an RNN is:
\begin{align}
    \textbf{h}_t & = \phi (\textbf{W}\textbf{x}_t + \textbf{U}\textbf{h}_{t-1} + \textbf{b}),
\end{align}
where $\textbf{W}$, $\textbf{U}$ are weight parameters, $\textbf{b}$ is the bias parameter, and $\phi(\cdot)$ is a nonlinear activation function.

Although the vanilla RNNs are useful, they suffer from the problem of exploding or vanishing gradients which makes them fail in capturing long-term dependencies effectively. The Gated Recurrent Unit (GRU) \cite{cho2014GRU} was proposed as a simplification of the Long-Short Term Memory (LSTM) \cite{hochreiter1997LSTM} module which overcomes the vanishing gradient problem in the simplest RNNs and has less parameters than LSTM. A typical GRU is formulated in the following way:
\begin{align}
\begin{split}
    \textbf{r}_t & = \sigma(\textbf{W}_r \textbf{x}_t + \textbf{U}_r \textbf{h}_{t-1}),\\ 
    \textbf{z}_t & = \sigma(\textbf{W}_z \textbf{x}_t + \textbf{U}_z \textbf{h}_{t-1}),\\
    \tilde{\textbf{h}}_t & = tanh(\textbf{W} \textbf{x}_t + \textbf{U}_z(\textbf{r}_t \odot  \textbf{h}_{t-1})),\\ 
    \textbf{h}_t & = \textbf{z}_t \odot \textbf{h}_{t-1} + (1-\textbf{z}_t) \odot \tilde{\textbf{h}}_t,
\end{split}
\end{align}
where $\sigma(\cdot)$ is the sigmoid function, $tanh(\cdot)$ is the hyperbolic tangent function, and $\odot$ is the element-wise multiplication operator.



\begin{figure*}[t]
\centering
\begin{minipage}[l]{1.9\columnwidth}
\centering
\includegraphics[width=0.95\textwidth]{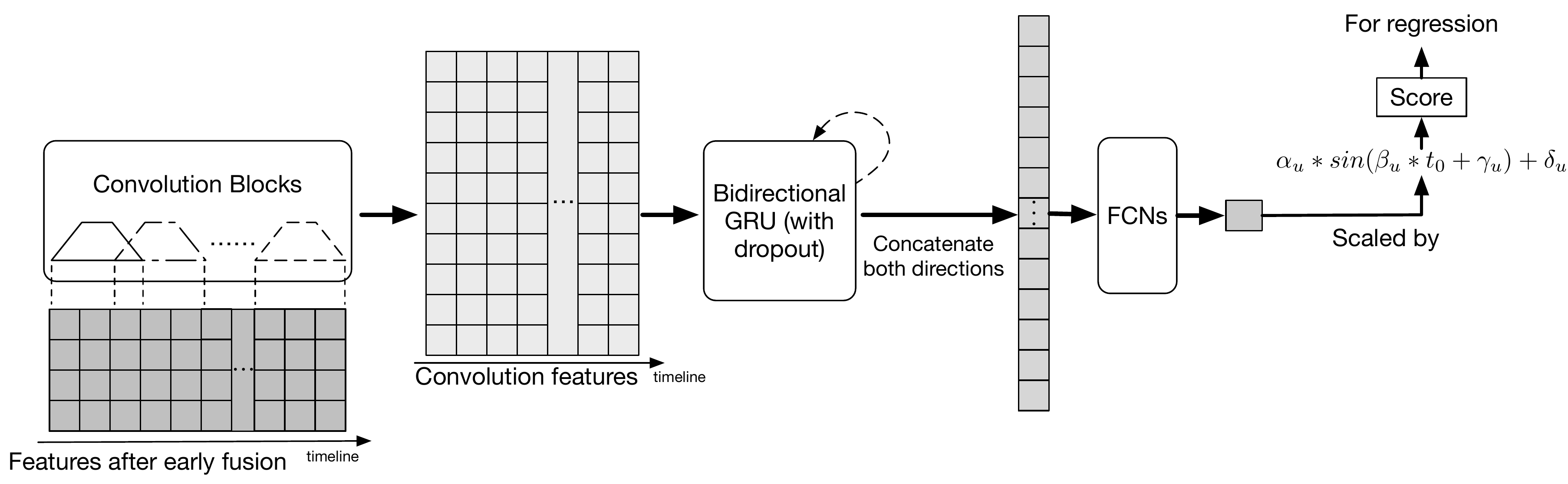}
\end{minipage}
\caption{The architecture of {\pro}.}
\label{fig:architecture}
\end{figure*}




\section{{\pro} Architecture}
In this section, we propose the {\pro} model that utilizes early-fused features, combines CNNs with RNNs, and considers personal circadian rhythm. The architecture of {\pro} is illustrated in Figure \ref{fig:architecture}. In following sections, we give detailed explanations of each component in {\pro}, i.e. early fusion of features, stacking CNNs and RNNs and considering each person's circadian rhythm.

\subsection{Early Fusion of Features}

Although DeepMood~\cite{cao2017deepmood} works well by using different networks to process features of different views and concatenate them in a later stage of the model (which we refer to as \emph{late fusion} approach), we instead propose to use \emph{early fusion} methods that align the alphanumeric keypresses with the accelerometer values before feeding them into any downstream machine learning models. The motivation behind \emph{early fusion} is that, aligning features of different views can provide extra information of the temporal relations between features.

\begin{figure}[!ht]
\centering
\subfigure[EF-fillna that fills the unaligned alphanumeric features with zeros]{
\begin{minipage}[l]{0.95\columnwidth}
\centering
\includegraphics[width=1\textwidth]{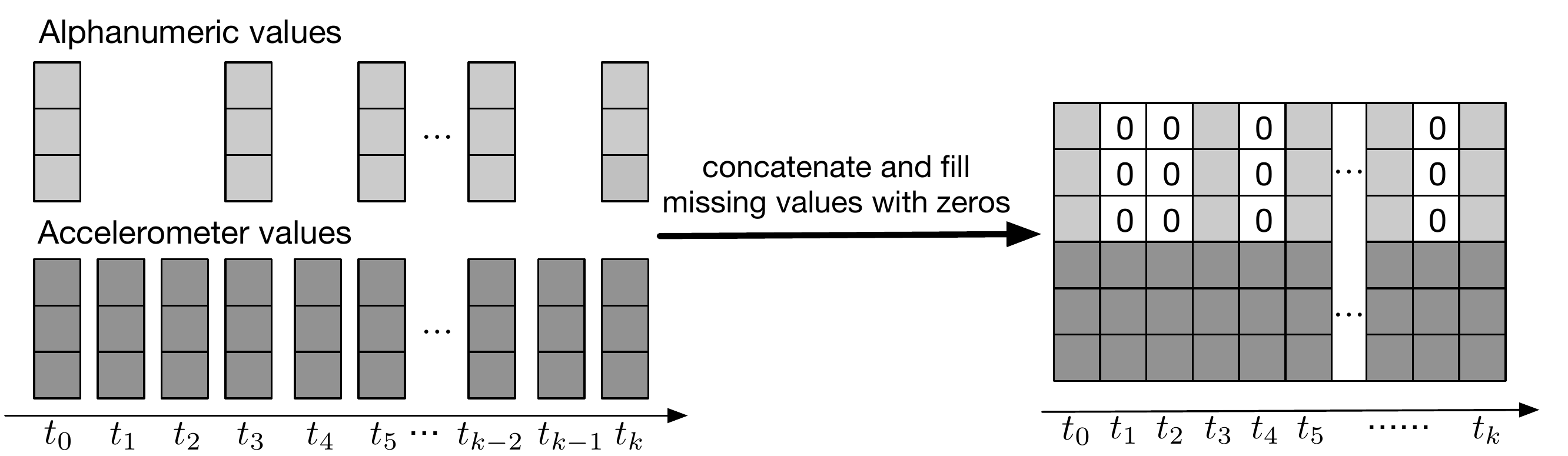}
\end{minipage}
\label{fig:early_fusion_long}
}\\
\subfigure[EF-dropna that discards all the unaligned accelerometer values]{
\begin{minipage}[l]{0.95\columnwidth}
\centering
\includegraphics[width=1\textwidth]{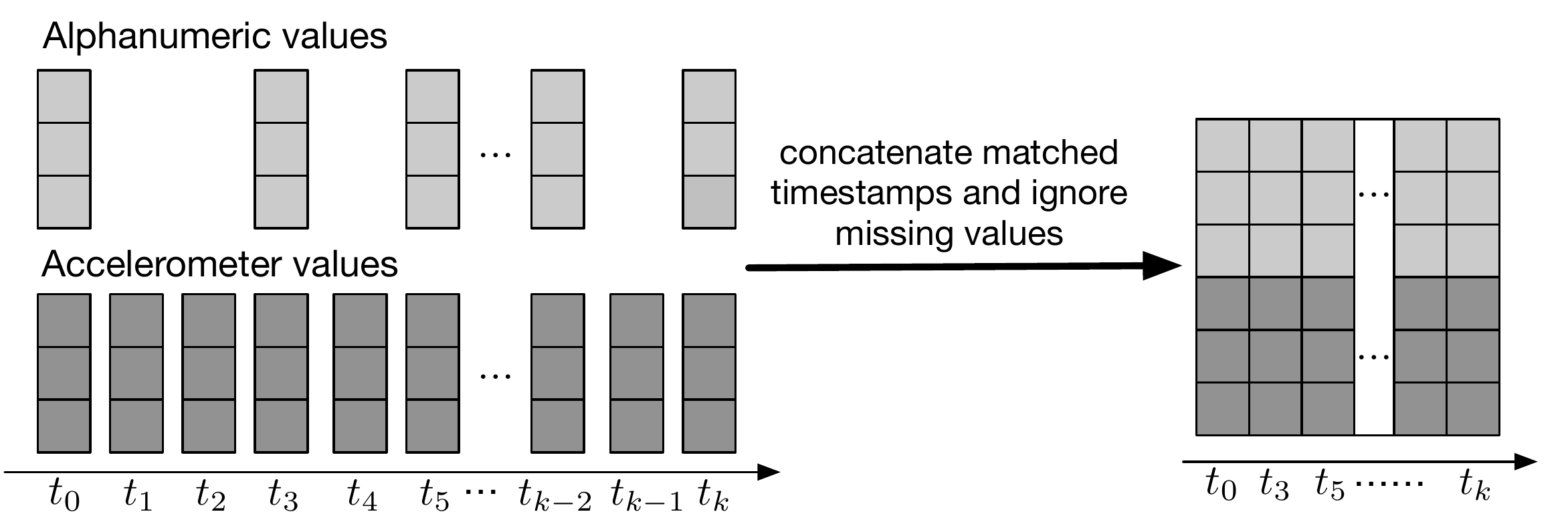}
\end{minipage}
\label{fig:early_fusion_short}
}
\caption{Two early fusion based methods. Both methods align the alphanumeric features to the nearest accelerometer features, but they deal with missing alphanumeric features in different manners.}
\label{fig:early_fusion}
\end{figure}

An intuitive early fusion approach, named as \emph{EF-dropna}, is to find the accelerometer value whose timestamp is the closest to each alphanumeric keypress, align them together, and drop the unaligned accelerometer values.
However, there are more accelerometer values than alphanumeric keypresses, and the abandon of those unaligned accelerometer values will certainly result in information loss. Therefore, another early fusion approach, named as \emph{EF-fillna}, is to retain the accelerometer sequence and fill the unaligned features in the alphanumeric sequence with zero values. These two early fusion methods are illustrated in Figure \ref{fig:early_fusion}. We try both early fusion methods in our proposed model  and name them \textbf{\pro-fillna} and \textbf{\pro-dropna} respectively.

\subsection{Exploiting Local Typing Dynamics with Convolutional Neural Networks}
In order to exploit the local typing patterns, we use 1-D convolution to learn the features of an input sequence. Specifically, the convolution block used in this paper is a convolution layer followed by batch-normalization and ReLU activation. 

One advantage of using multi-layer CNNs to extract useful features from the sequence data instead of directly applying multi-layer RNNs is that CNN-based models can be trained much faster than RNNs without any sacrifice in prediction performance which will be validated in the experiments.

\subsection{Combining CNNs and RNNs}
\label{sec:cnn+rnn}
Using CNNs alone will lose the long-term temporal dynamics, since the kernels in CNNs are designed to capture only local features within a small window. On the contrary, although RNNs take more time for training, they can produce features that capture the overall dynamics of the input sequence. Therefore, we propose to use a combination of CNNs and RNNs, so that we can make use of both their advantages in learning local patterns and temporal dependencies.

To do so, we first feed features into two convolution block and then into a bi-directional GRU module. Specifically, we take the output features from the second convolution block, split the features along the temporal dimension, and feed the split features into the GRU sequentially. We concatenate the last output features from the GRU in two directions to form a single feature vector. Also, since we apply \emph{early fusion} to the features, we only have one pathway and do not need to concatenate results from different views.

\subsection{Periodic Dynamics and Personalized Mood Prediction}
A circadian rhythm is any biological process that displays an endogenous oscillation of about 24 hours \cite{edgar2012peroxiredoxins}. It is running in the background of human brain and cycles between sleepiness and alertness at regular intervals. It is commonly known that individual depressive moods vary according to the circadian rhythm, as well as the day of the week \cite{suhara2017deepmood}.

In order to exploit the circadian rhythm and other periodic patterns as discussed in Section \ref{sec:data_time}, as a rough approximation, we propose to use the $Sine$ function for time-based calibration of the final prediction.
Specifically, suppose the output of the last fully-connect layer is $x$, before producing the final score $s$ for regression, we scale $x$ by the value of the $Sine$ function which takes the current time as input. 


Furthermore, smart-phone users usually have very distinct typing dynamics. As shown in Section \ref{sec:data_subject}, each subject may have different baselines in terms of mood states, even with similar typing dynamics. Therefore, we should make personalized mood prediction rather than using a subject-unaware model. This could enable us to improve prevention and treatment outcomes by better incorporating individual patient characteristics.

In order to provide personalized mood prediction, the final prediction should be further adjusted per subject. Hence, it is more intuitive to use a different set of parameters $\{\alpha, \beta, \gamma, \delta\}$ for each person, which are learned automatically by gradient descent and back-propagation. The calibration for user $u$ is given by:
\begin{align}
    s = x * (\alpha_u * sin (\beta_u * t_0 + \gamma_u) + \delta_u),
\end{align}
where $t_0$ is the starting time of the input sequence ({e.g.} a phone usage session), represented by the number of hours passed since the start of the earliest session of the subject.

There are many choices of periodic functions, but since any periodic function can be approximated by a Fourier series which is made of many $Sine$ functions, here we use only one $Sine$ function and demonstrate its effectiveness in helping mood prediction.


\section{Experiments}
In this section, we evaluate the proposed \pro \ model HDRS and YMRS regression tasks, study the convergence efficiency of different methods and investigate the effects of changing the size of training set on regression tasks.


\begin{table}[!ht]
\centering
\caption{Configuration of hyper-parameters.}
\label{tab:para_cofig}
\begin{tabular}{|l|c|}
\hline
Parameter            & Value \\ \hline \hline
learning rate        & 0.001 \\ \hline
batch size           & 256   \\ \hline
training epochs      & 200   \\ \hline
dropout ratio        & 0.1   \\ \hline
min sequence length  & 10    \\ \hline
max sequence length  & 100   \\ \hline
GRU hidden dimension & 20    \\ \hline
\end{tabular}
\end{table}

\subsection{Experimental Setup}
\label{sec:exp_setup}

For the prediction of depression score HDRS and mania score YMRS, we treat them as  regression tasks where HDRS/YMRS scores are used as labels. \emph{Root-mean-square error} (RMSE) is used as the metric for both tasks. 
The features used in all the compared models are defined below:
\begin{itemize}
\item \textbf{Alphanumeric sequence}: The features in the alphanumeric sequence are represented by a 4-dimensional vector that includes {\em duration of a keypress}, {\em time since last keypress}, {\em horizontal distance to last keypress}, and {\em vertical distance to last keypress}.
\item \textbf{Accelerometer sequence}: The features in the accelerometer sequence are represented by a 3-dimensional vector where each dimension represents the accelerometer values along X, Y and Z axises, respectively.
\end{itemize}

Each session is composed of these two sequences, a alphanumeric sequence and a accelerometer value sequence. We truncate sessions that contain more than 100 alphanumeric characters, and remove sessions if they contain less than 10 alphanumeric characters. This leaves us with 14,613 total sessions. For each subject, we use the earliest 80\% of sessions for training and the rest for testing.

The model is implemented in PyTorch, and runs on a Ubuntu system with an NVIDIA Titan X Pascal GPU. The hyper-parameters for the proposed model and all baselines are fixed to the same, and RMSProp \cite{hinton2012rmsprop} is chosen as the optimizer. The two convolutional blocks of \pro\ have 10 and 20 kernels respectively, and each kernel is of size 3 and stride 2. Note the CNN baseline has a third convolutional block with 30 kernels, each kernel if of size 3 and stride 2, while \pro \ uses only two convolution blocks in order to preserve enough information on the temporal dimension for GRUs. Other hyper-parameter values are listed in Table \ref{tab:para_cofig}. Our code is open-source at \textbf{\url{https://github.com/stevehuanghe/dpMood}}.

\subsection{Baseline Methods}
Since Cao et al.~\cite{cao2017deepmood} have shown that deep architectures work better than traditional methods in mood prediction, we only compare with baselines that use deep neural networks. The compared methods are introduced as follows:
\begin{itemize}
\item RNN (DeepMood): A model that feeds each sequence to a separate bi-directional GRU, and the concatenated outputs are connected to a fully-connected network for regression. It is the same as the deep architecture proposed in \textbf{DeepMood}~\cite{cao2017deepmood} that uses {\em alphanumeric characters}, {\em accelerometer values} and {\em special characters} as features, and we re-implement it in PyTorch. 
\item CNN: A CNN-based model which stacks three convolutional blocks followed by a max-pooling layer that reduces the number of channels to 1. Late fusion is applied to compare with DeepMood, which means there is a separate convolutional neural network for each kind of input features. The resulting features of each network are then concatenated into a single feature which is then put into a fully-connected network to produce a single scalar for regression.
\item CNNRNN: A model that stacks CNN and RNN together as in \pro, uses \emph{late fusion}, without any time-based or personalized calibration.
\item CNNRNN-Cr: The CNNRNN model that learns the same circadian for all users, {\em i.e.}, no personalization.
\item CNNRNN-PsCr: The CNNRNN model that explores each person's circadian rhythm.
\item CNNRNN-fillna: The CNNRNN model that uses early fusion method  \emph{EF-fillna}.
\item CNNRNN-dropna: The CNNRNN model that uses early fusion method  \emph{EF-dropna}.
\item \textbf{\pro-fillna}: The proposed \pro \ model which stacks CNN with RNN, learns personal circadian rhythms and utilizes the early fusion method \emph{EF-fillna}
\item \textbf{\pro-dropna}: The proposed \pro \ model which stacks CNN with RNN, learns personal circadian rhythms and utilizes the early fusion method \emph{EF-dropna}
\end{itemize}

\begin{table*}[!htb]
\centering
\caption{Regression performance on different tasks, shown in ``{mean ($\pm$ std)}'' format, evaluated using RMSE. There are two sets of experiments, one with both bipolar and control subjects (w/ ctrl), the other with only bipolar subjects (w/o ctrl).}
\label{tab:results}
\begin{tabular}{|l|c|c|c|c|}
\hline
Model                  & HDRS w/ ctrl                 & YMRS w/ ctrl                 & HDRS w/o ctrl                & YMRS w/o ctrl                \\ \hline \hline
RNN                    & 5.410 ($\pm$ 0.054)          & 3.700 ($\pm$ 0.034)          & 4.765 ($\pm$ 0.048)          & 4.150 ($\pm$ 0.097)          \\ \hline
CNN                    & 5.077 ($\pm$ 0.119)          & 3.600 ($\pm$ 0.050)          & 4.806 ($\pm$ 0.057)          & 4.085 ($\pm$ 0.048)          \\ \hline
CNNRNN                 & 4.671 ($\pm$ 0.097)          & 3.477 ($\pm$ 0.042)          & 4.526 ($\pm$ 0.067)          & 4.129 ($\pm$ 0.110)          \\ \hline
CNNRNN-Cr              & 7.023 ($\pm$ 0.043)          & 4.032 ($\pm$ 0.034)          & 8.647 ($\pm$ 0.031)          & 5.056 ($\pm$ 0.049)          \\ \hline
CNNRNN-PsCr            & 2.818 ($\pm$ 1.439)          & 3.202 ($\pm$ 0.243)          & 5.164 ($\pm$ 1.686)          & 4.103 ($\pm$ 0.398)          \\ \hline
CNNRNN-fillna          & 6.304 ($\pm$ 0.105)          & 3.798 ($\pm$ 0.027)          & 5.165 ($\pm$ 0.067)          & 4.198 ($\pm$ 0.021)          \\ \hline
CNNRNN-dropna          & 6.698 ($\pm$ 0.063)          & 6.698 ($\pm$ 0.063)          & 5.196 ($\pm$ 0.023)          & 4.312 ($\pm$ 0.032)          \\ \hline
\textbf{dpMood-fillna} & \textbf{2.400} ($\pm$ 1.281) & \textbf{3.104} ($\pm$ 0.422) & \textbf{3.064} ($\pm$ 1.381) & \textbf{4.013} ($\pm$ 0.294) \\ \hline
\textbf{dpMood-dropna} & \textbf{2.376} ($\pm$ 1.065) & \textbf{3.020} ($\pm$ 0.254) & \textbf{3.641} ($\pm$ 1.880) & \textbf{3.921} ($\pm$ 0.175) \\ \hline
\end{tabular}
\end{table*}

\subsection{Regression Performance}
Here we conduct two sets of experiments, one with all 20 subjects, and the other with only 12 bipolar subjects (without control subjects). We did not use the same random seed for all models, and let the compared models run 20 times (200 epochs for each run) so that we can calculate their means and standard deviations. The results are shown in Table \ref{tab:results}. Overall, our \pro\ model achieves the lowest average RMSE among all compared methods. By comparing the results of RNN and CNN, we can see that the performance of fully-convolutional model is comparable with the performance of RNN model, even if CNN fails to capture the long-term dependencies of the sequences. This reveals that the local patterns in the typing dynamics that are captured by the CNN model is as important as the temporal dependencies learned by the RNN model.

When we combine the CNN and RNN model, the CNNRNN model achieves lower regression error than the separate CNN and RNN models by an average margin of 7\% for the HDRS (with controls), and 6\% improvement for YMRS (with controls). For HSRS regression without controls, the CNNRNN model performs slightly better than CNN and RNN, while in the YMRS regression without controls, CNNRNN is about the same as CNN and RNN, but with a larger variance in RMSE. Overall, the performance of the CNNRNN model indicates that preserving both local and global typing dynamics can help mood predictions.

As for the CNNRNN-Cr model, we can see that adding the same calibration function to all subjects does not improve the performance, and it even leads to a large increase in RMSE when compared to  the simple CNN and RNN model, which may be because of the fact that different people have very different personal traits, especially for the bipolar subjects, as shown in Section \ref{sec:data}.

When learning a different calibration function for each subject, the CNNRNN-PsCr model is able to achieve better performance than all previously mentioned methods in both HDRS and YMRS regression with controls, which shows the potential of considering personal circadian rhythms in mood prediction. However, CNNRNN-PsCr is worse than other methods on tasks that do not have control subjects, which may be because the patterns of normal persons are easier to learn than that of bipolar subjects. 

From the performance of CNNRNN-fillna and CNNRNN-dropna, we can see that using early fusion alone does not help better predicting HDRS or YMRs, and they are slightly inferior to the CNNRNN model that uses late fusion.

As for our \pro-fillna and \pro-dropna, they both perform better than all baselines and have lower average RMSEs, which proves the effectiveness of considering personal circadian rhythms and using early fusion.
By comparing our proposed models with CNNRNN-PsCr, we can see that early fusion can help learning better personal traits, while the results of CNNRNN-fillna and CNNRNN-dropna show that early fusion works better when considering circadian rhythms.

\begin{figure}[!htb]
\centering
\begin{minipage}[l]{1\columnwidth}
\centering
\includegraphics[width=0.95\textwidth]{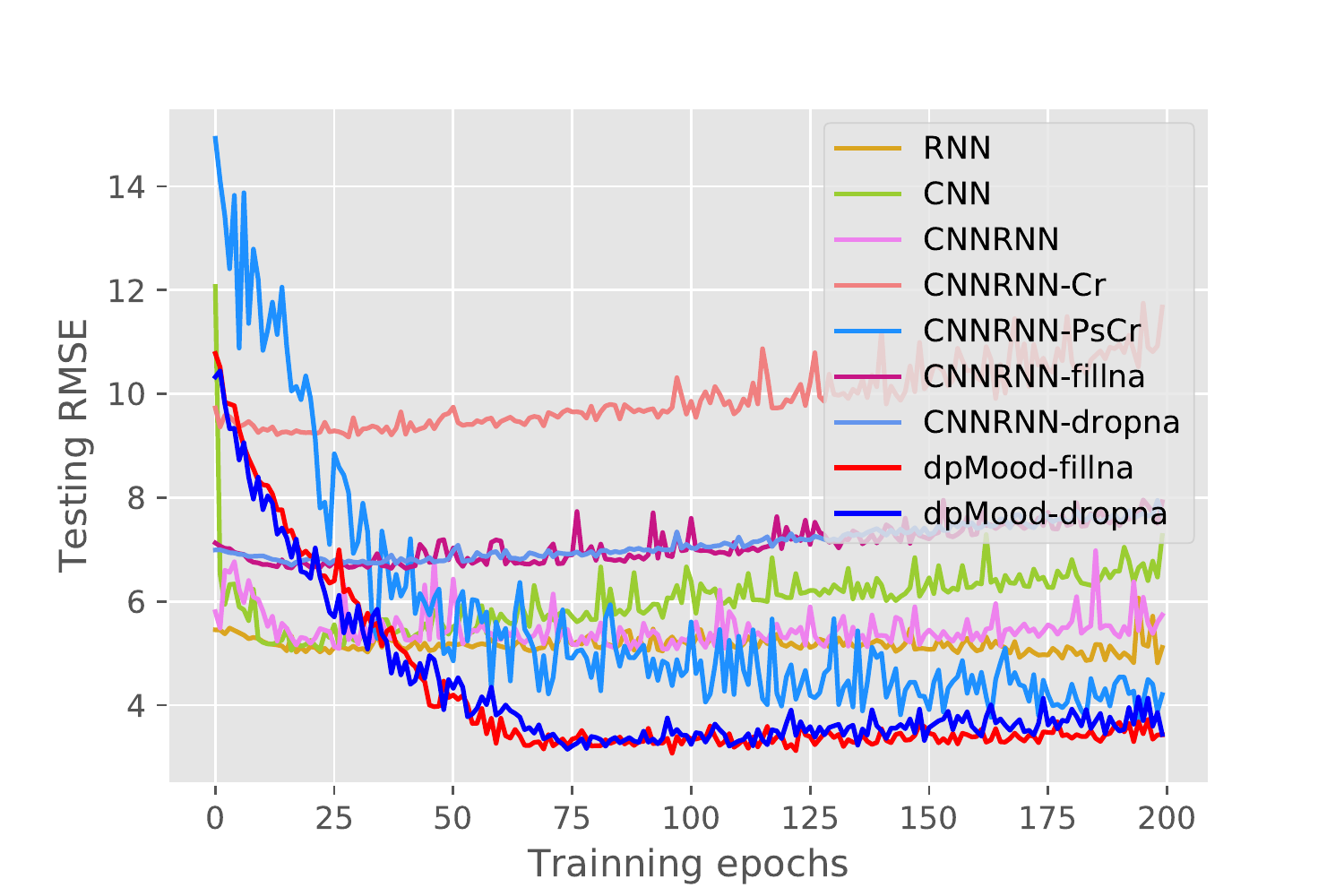}
\end{minipage}
\caption{Convergence curve of HDRS regression with controls.}
\label{fig:converge_cmp}
\end{figure}

\subsection{Convergence Efficiency}

In this section, we analyze the convergence of \pro \ and the baseline methods. The random seed is fixed as 1234 for all models, and each model is trained for 200 epochs and is evaluated on the test set after each epoch. As shown in Figure \ref{fig:converge_cmp}, although our \pro-fillna and \pro-dropna converge not as quickly as some of the baselines,  \pro-fillna achieves it best performance at the 97th epoch, while \pro-dropna takes 75 epochs, and \pro-fillna has slightly lower RMSE than \pro-dropna. The RNN model converges slowly. After a fast decrease in test RMSE in the first 30 epochs, its convergence procedure slows down as training goes on. In the experiment with 500 epochs, the RNN model achieves its best test RMSE at around the 480$th$ epoch. As for the CNN model, we can see that it converges very quickly and reaches its lowest RMSE at the 17$th$ epoch, after which it starts to overfit with a decrease in the performance. 

Compared to the CNN model, the CNN+RNN model takes more epochs to converge to its best performance (88 epochs), which is normal because it contains an RNN that requires longer time to converge. However, because of the CNN part in the model that significantly reduces the length of input features to the RNN part, the convergence efficiency of the CNNRNN model is still much better than the RNN model. The CNNRNN model takes about 76 epochs to reach its best test RMSE. 

Clearly that the CNNRNN-Cr model performs the worst, which indicates that it is inappropriate to treat all users' circadian rhythms as they are the same. The CNNRNN-PsCr model, with better performance than all previously mentioned baselines, also preserves a fast convergence rate. In the experiments, it reaches the lowest test RMSE at the 195$th$ epoch which is slower than the CNNRNN model, but it outperforms CNNRNN by 40\% on HDRS regression task with control subjects.

\subsection{Sensitivity Regarding the Size of Training Data}

In this section, we analyze the performance of our method with respect to the size of training data. We fix the random seed as 1234 and train each model for 200 epochs. We start with using only 30\% of the whole data as training and use the rest for testing, and gradually increase the portion of training data up to 70\%. This experiment is conducted on HDRS regression task with all 20 subjects, and the results are shown in Table \ref{tab:ratio}. As we can see, all models benefit from the increase of training data, but our \pro-fillna achieves the lowest regression error with only 30\% of the whole data as training data, which indicates that our model is able to work well with small training data and that it is important to consider each user's personal trait. As for our \pro-dropna, although it has slightly higher RMSE than CNNRNN-PsCr when trained on small training sets, it outperforms CNNRNN-PsCr when we use 70\% of the whole data for training.

\begin{table}[!htb]
\centering
\caption{RMSE of HDRS regression with respect to different train/test splitting ratio, with control subjects}
\label{tab:ratio}
\begin{tabular}{|l|c|c|c|c|c|}
\hline
Model/Train ratio         & 30\%           & 40\%           & 50\%           & 60\%           & 70\%           \\ \hline
RNN                       & 6.159          & 6.073          & 6.080          & 5.944          & 5.893          \\ \hline
CNN                       & 6.170          & 6.092          & 5.906          & 5.840          & 5.701          \\ \hline
CNNRNN                   & 5.668          & 5.456          & 5.398          & 5.307          & 5.292          \\ \hline
CNNRNN-Cr              & 8.240          & 8.126          & 8.037          & 8.011          & 8.010          \\ \hline
CNNRNN-PsCr            & 3.777 & {3.373} & {3.140} & {2.979} & {2.781} \\ \hline
\textbf{dpMood-fillna} & \textbf{3.297} & \textbf{3.195} & \textbf{3.101} & \textbf{3.017} & \textbf{2.601} \\ \hline
\textbf{dpMood-dropna} & \textbf{3.877} & \textbf{3.488} & \textbf{3.167} & \textbf{3.166} & \textbf{2.596} \\ \hline
\end{tabular}
\end{table}

\subsection{Visualizing Personalized Calibration Functions}

\begin{figure}[!thb]
\centering
\subfigure[]{
\begin{minipage}[l]{1\columnwidth}
\centering
\includegraphics[width=0.9\textwidth]{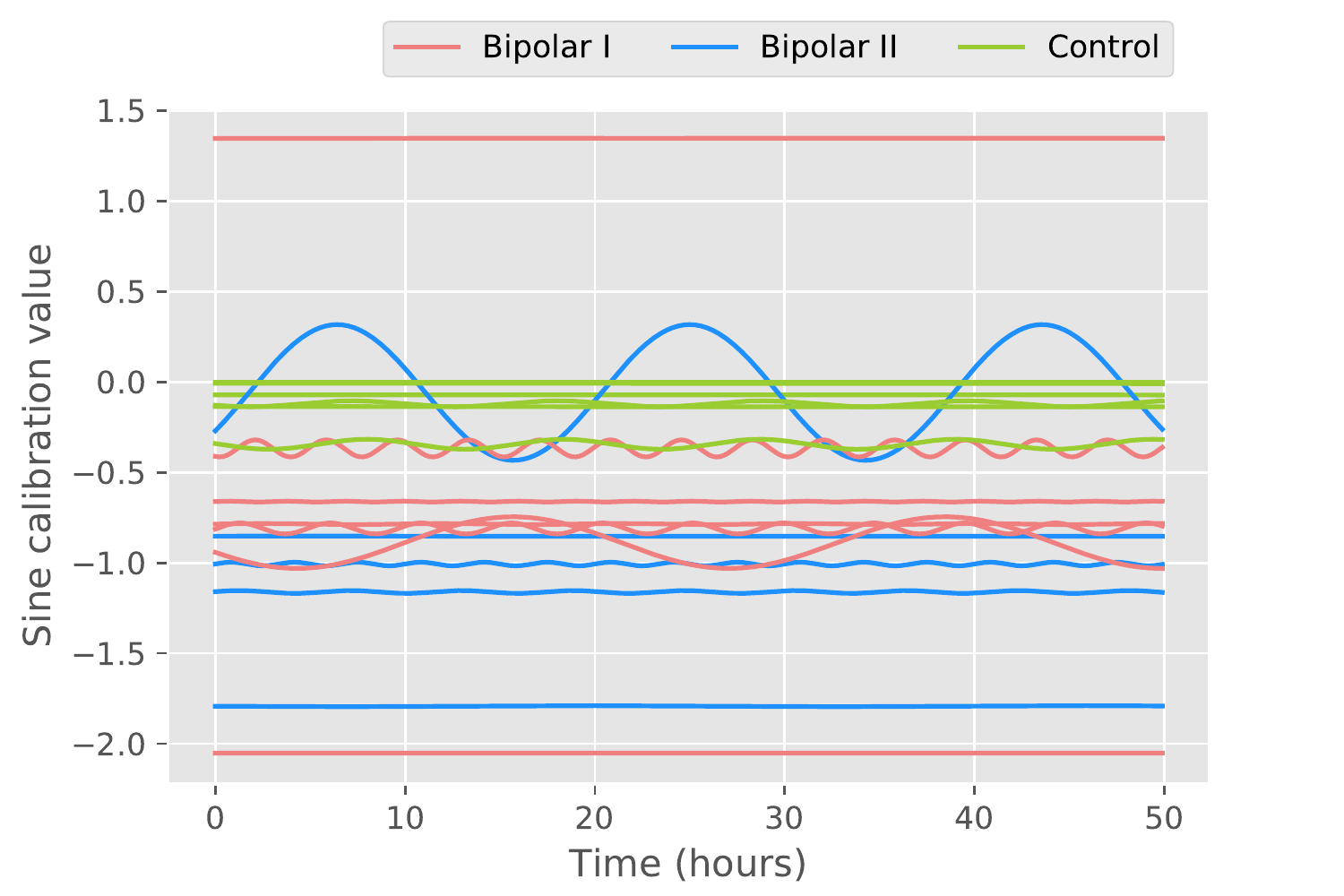}
\label{fig:sine_hdrs}
\end{minipage}
}
\subfigure[]{
\begin{minipage}[t]{1\columnwidth}
\centering
\includegraphics[width=0.9\linewidth]{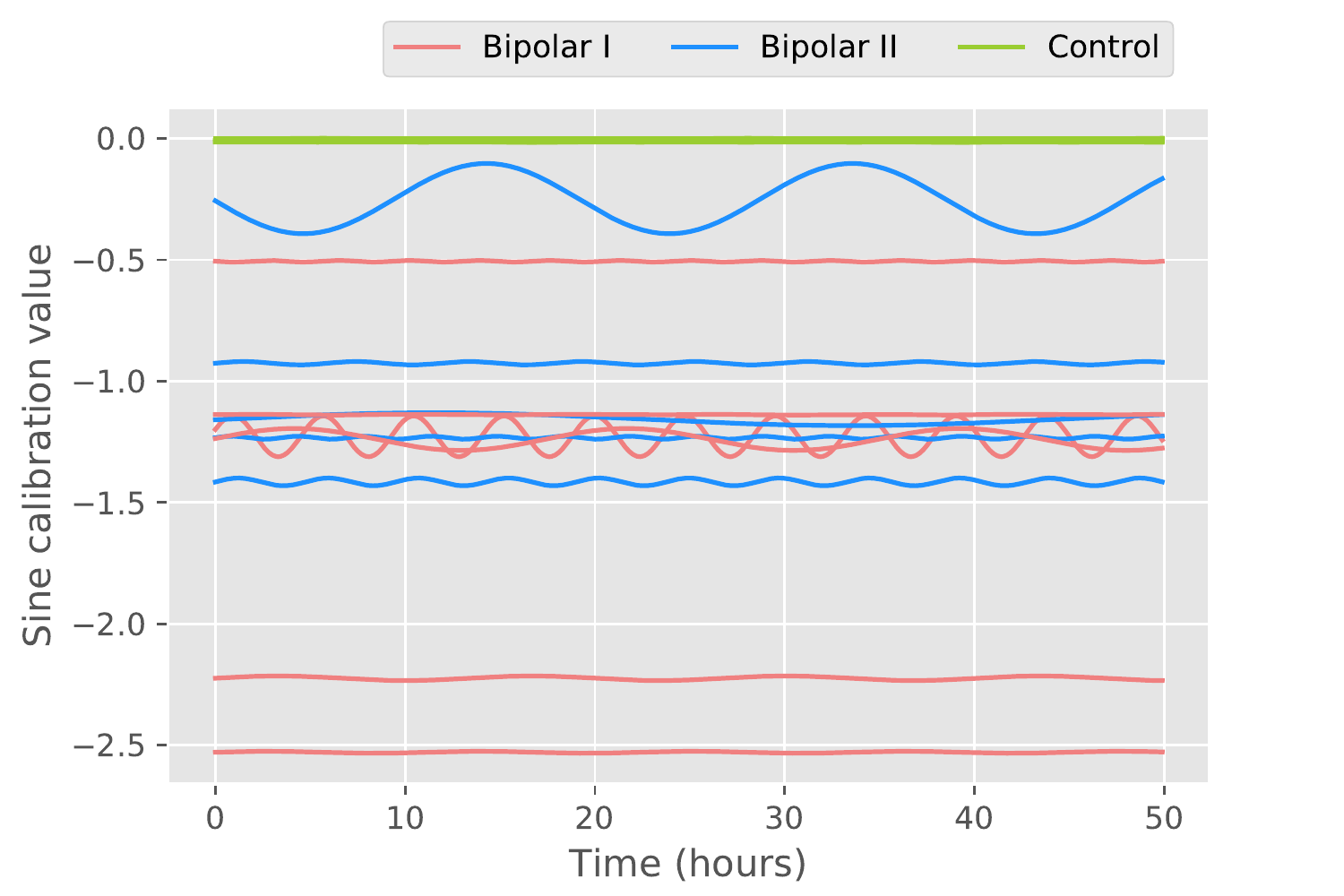}
\label{fig:sine_ymrs}
\end{minipage}
}
\caption{Visualization of each patient's calibration function: (a) HDRS; (b) YMRS.}
\label{fig:sine_vis}
\end{figure}

In this section, we visualize the learned calibration for each subject and analyze the differences between subjects. Figure \ref{fig:sine_hdrs} shows the $Sine$ functions learned for the HDRS regression task, while Figure \ref{fig:sine_ymrs} shows the functions learned for the YMRS regression task. We can see that most subjects have very distinct $Sine$ functions, which indicates that every subject has her own circadian rhythm, and some subjects have longer periods than the others.

The colors in Figure \ref{fig:sine_vis} indicate the diagnosis of each subject, where the \emph{control} group contains subjects who are diagnosed with neither \emph{bipolar I} or \emph{bipolar II}, and we can consider that \emph{bipolar I} is more severe than  \emph{bipolar II}.
An interesting discovery is that the offset terms $\delta_u$ are effectively clustered {\em w.r.t.} the diagnosis. 
Figure \ref{fig:sine_hdrs} shows that the \emph{control} group has positive offsets, and subjects in the \emph{bipolar I} and \emph{bipolar II} group usually have a negative offset.
From Figure \ref{fig:sine_ymrs} we can see that the \emph{control} group has calibration values that are very close to 0, which corresponds to the fact that a large portion of the YMRS scores for subjects in the \emph{control} group are 0.





\section{Related Work}

Mood prediction has been explored in different manners such as using smart-phones and wearable devices to record personal traits like sleeping \cite{sano2015recognizing}, voice acoustic \cite{lu2012stresssense}, and social patterns \cite{moturu2011socialsense,ma2012daily}. Recently, Suhara et al. \cite{suhara2017deepmood} design a method that uses peoples' self-reported mood histories to predict depression. However, their method highly relies on the quality of users' self-assessment, which may be additional burden to users and they may stop reporting after a while. Cao et al.~\cite{cao2017deepmood} use keyboard typing as features and professional medical diagnosis as labels for predicting depression and mania, but their model does not capture the local patterns in the typing dynamics or consider each person's circadian rhythm.

The task of mood prediction is closely related to supervised sequence prediction. A brief survey by Xing et al. \cite{xing2010brief} categorizes sequence data into five groups: simple symbolic sequences, complex symbolic sequences, simple time series, multivariate time series, and complex event sequences. Models used for sequence classification are also categorized into three groups: feature based methods \cite{lesh1999mining,aggarwal2002effective,leslie2004fast,ji2007mining,ye2009time}, sequence distance based methods \cite{keogh2000scaling,keogh2003need,ratanamahatana2004making,wei2006semi,xi2006fast,ding2008querying,lodhi2002text,she2003frequent,sonnenburg2005large}, and model based methods \cite{cheng2005protein,yakhnenko2005discriminatively,srivastava2007hmm}. This work is related to the feature based approach, but we use deep learning models to learn higher level features for regression tasks.


This work is also related to sentence classification in natural language processing \cite{kim2014convolutional,zhang2015character}. The common part of sentence classification and mood prediction is that both tasks use sequence data as input. Text sequences and time series data are similar to each other in that the order of elements in a sequence is important to the meaning of the sequence. Both text sequence and typing sequence have local patterns. For text, several words together form an n-gram phase that represents a certain meaningful concept. In our case, although we do not know the meaning of each keypress, the local patterns can still represent some information similar to n-grams, since the keypresses are mostly alphanumeric characters. However, the sentence classification and mood prediction tasks are different from some perspectives. In sentence classification, each word is represented by an embedding vector that indicates its position in the latent space, while in mood prediction, each feature is obtained from the real-world sensors thereby having certain physical meaning. Although we also use CNNs as in \cite{kim2014convolutional,zhang2015character,kalchbrenner2014convolutional}, we need to deal with data from multiple views, while in sentence classification, the data usually come from a single view.

\section{Conclusion}
This paper studies the problem of mood prediction using typing dynamics collected from smart-phones, and proposes an end-to-end deep architecture that incorporates both CNNs and RNNs. Moreover, the proposed \pro \ model considers each person's circadian rhythm and adjusts the predictions accordingly. Extensive experiments demonstrate the power of using the combination of CNNs and RNNs in mood prediction, and that modeling each person's circadian rhythm is critical for achieving more accurate predictions. In addition, we study the effect of early fusion for multi-view sequence data and compare it with late fusion, and find that early fusion help improve the performance of our \pro\ model in the given tasks.

The Precision Medicine Initiative\footnote{https://allofus.nih.gov} is a recent project that aims to improve prevention and treatment outcomes by better incorporating individual patient characteristics, and mobile technologies including smart-phones and wearable devices are expected to play a significant role in these efforts. This work demonstrates the feasibility and potential of such efforts. 




\section*{Acknowledgment}
This work is supported in part by NSF through grants IIS-1526499, IIS-1763325, and CNS-1626432, and NSFC 61672313.

\bibliographystyle{IEEEtran}
\bibliography{bibliography} 

\end{document}